\documentclass[11pt]{article}

\usepackage{amsfonts,amsmath,amssymb,amscd,amsthm}
\usepackage{enumitem}
\usepackage{mathrsfs}

\def\p{\partial}
\usepackage{epsfig,color}
\usepackage{color}
\usepackage{multirow}
\usepackage{epstopdf}
\usepackage{color,graphicx}
\usepackage{float}
\usepackage{subfigure}
\usepackage{appendix}
\usepackage{natbib}
\usepackage{indentfirst}

\newtheorem{thm}{Theorem}[section]
\newtheorem{lemma}[thm]{Lemma}
\newtheorem{corollary}[thm]{Corollary}

\newtheorem{exam}[thm]{\noindent Example}
\newtheorem{assum}[thm]{Assumption}

\newtheorem{remark}[thm]{Remark}

\numberwithin{equation}{section}

\begin{document}

\title{Global Closed-form Approximation of Free Boundary for Optimal Investment Stopping Problems}
\vskip 4mm

\author{Jingtang Ma\thanks{School of Economic Mathematics,
Southwestern University of Finance and Economics, Chengdu,
611130, P.R. China (Email: mjt@swufe.edu.cn).
The work
was supported by National Natural Science Foundation of
China (Grant No. 11671323), Program for New Century Excellent Talents in University (China Grant No. NCET-12-0922)  and the Fundamental Research Funds for the Central Universities, P.R. China (JBK1805001).}, Jie Xing\thanks{School of Economic Mathematics,
Southwestern University of Finance and Economics, Chengdu,
611130, P.R. China (Email: 1160202Z1010@2016.swufe.edu.cn).} and  Harry Zheng
\thanks{Corresponding author. Department of Mathematics, Imperial College, London SW7 2BZ, UK
(Email: h.zheng@imperial.ac.uk).}}

\date{}
\maketitle

\begin{abstract}
In this paper we study a utility maximization problem with both optimal control and optimal stopping in a finite time horizon. The value function can be characterized by a variational equation that involves
a free boundary problem of a fully nonlinear partial differential equation. Using the dual control method, we derive the asymptotic properties of the dual value function and the associated dual free boundary for a class of utility functions, including power and non-HARA utilities. We construct a global closed-form approximation to the dual free boundary, which greatly reduces the computational cost.
 Using the duality relation, we find the approximate formulas for the optimal value function, trading strategy,  and exercise boundary for the optimal investment stopping problem. Numerical examples show the approximation is robust, accurate and fast.
\end{abstract}

\vspace{0.5cm}


\vspace{0.5cm}

\section{Introduction}

There has been extensive research in utility maximization. Two main approaches are stochastic control (dynamic programming, HJB equation) and convex duality (static optimization, martingale representation). For excellent expositions of these two methods in utility maximization,
see \cite{fs1993},  \cite{KARATZAS1998}, \cite{PHAM2009}, and the references therein.

A variant of utility maximization of terminal wealth is that   investors may stop the investment before or at the maturity  to achieve the overall maximum of the expected utility, which naturally leads to
 a mixed optimal control and stopping problem. The early work on this line includes
\cite{KARATZAS2000} and \cite{DAYANIK2003}  for properties of the value function at the initial time, \cite{CECI2004}  for existence of viscosity solution of the variational equation, \cite{HENDERSON2008}  for equivalence of the value function in the presence of a Markov chain process and power utility. None of the above papers discusses the free boundary problem.
  \cite{JIAN2014} apply the dual transformation method to convert the nonlinear variational equation with power utility into an equivalent
 free boundary problem of a linear PDE and analyse qualitatively the properties of the free boundary and optimal strategies. The work is further extended in \cite{GUAN2017} to a problem with a call option type terminal payoff and power utility.

It is well known that finding the free boundary of a variational equation is a difficult problem, see \cite{peskir2006}.   One good example is American options pricing problem. The free boundary separates the exercise region from the continuation  region and satisfies an integral equation which can be hardly solved, see \cite{detemple2005}.  Finding the free boundary is much more difficult for  the optimal investment stopping problem  than    for the  American options pricing problem as
   the former has a nonlinear PDE in the continuation region and a non-Lipschitz continuous utility function and  may have one or more free boundaries  whereas the latter has a linear PDE in the continuation region and a Lipschitz continuous payoff function and a unique free boundary. The dual transformation  in \cite{JIAN2014} and \cite{GUAN2017} is a step in the right direction to simplify the primal nonlinear variational equation into the dual linear variational equation, however, finding the free boundary  remains a challenging and  open problem.

In this paper we study an optimal investment stopping problem for general utility functions with a requirement that the wealth is above a threshold value which could be a liability or the minimum living standard,  called portfolio insurance. Using the dual transformation approach as in  \cite{JIAN2014} and \cite{GUAN2017}, we convert the primal variational equation into an equivalent free boundary problem of a linear PDE  and show there exists a unique smooth free boundary that satisfies some integral equation for a class of utility functions, including power and non-HARA utilities, see Theorems \ref{Theorem-5.2} and \ref{Theorem 3.1}. We then apply the asymptotic analysis to characterize the limiting behaviour of the free boundary as time to maturity tends to zero and to infinite, see Theorems \ref{Theorem 3.2} and \ref{Theorem 4.4}. We construct a  simple function that has the same property as the free boundary with matched limiting behaviour and use it as a global closed-form approximation  to the free boundary, which is inspired by \cite{Xie2014} for a mortgage payment problem with a simple time-only, state-independent  payoff and known initial value of the process, in contrast to our non-Lipschitz state-dependent payoff and unknown initial value of the dual process.  Finally, using the duality relation, we recover the  primal value function and the corresponding free boundary,  see Theorem \ref{Theorem 5.1}.

The main contribution of this paper is that we give a global closed-form approximation (GCA) to the free boundary of an  optimal investment stopping problem for a class of general utility functions.  There are several decisive benefits of the GCA: it provides a simple analytic formula for separating the stopping region and the continuation region, it gives the dual value function a semi closed-form integral representation which makes possible finding the  optimal trading strategy in the continuation region, and it leads to fast and efficient computation. The key to this success is the explicit characterization  of the asymptotic properties of the free boundary for the dual optimal stopping problem.    To the best knowledge of the authors, this is the first time such results are reported in the literature for optimal investment stopping problems. Numerical tests show that GCA is accurate and  fast, compared with the binomial tree method which itself is practical and efficient in solving optimal investment stopping problems, see Example \ref{ex-6.2}.

The remaining of the paper is organized as follows. In Section \ref{sec:problem}, we introduce the optimal investment stopping problem, convert the HJB variational equation into an equivalent  dual variational equation, show the existence and uniqueness of the dual solution and its properties, and establish the corresponding results for the original problem.   In Section \ref{sec:formulas}, we present the main results of this paper  for a class of utilities which include power and non-HARA utilities, Theorem \ref{Theorem 3.1} shows the free boundary is monotone and  smooth and satisfies an integral equation, Theorems \ref{Theorem 3.2} and \ref{Theorem 4.4} characterize the asymptotic behaviour of the free boundary when time to maturity is close to zero or infinite,  Theorem \ref{Theorem 5.1} constructs a GCA to the free boundary.
We also give two examples  (Examples \ref{exm-nonhara} and  \ref{exam-K=0}) to illustrate the fundamental difference of utility maximization with portfolio insurance and without.
In Section \ref{sec:examples}, we  perform some numerical tests to compare the results derived with the GCA and with the binomial tree method and show the suggested GCA is accurate, fast, and robust.
In Section \ref{proofs}, we give the proofs of the main results. Section \ref{sec:conclusion} concludes.
 The appendix provides the proof of Theorem \ref{Theorem 4.1}
for the convenience of the reader.

\section{Optimal Investment Stopping Problems}\label{sec:problem}
We consider a complete market equipped with a probability space $(\Omega,\mathscr{F},P)$ together with a  natural filtration $(\mathscr{F}_t)$ generated by a standard Brownian motion $W$,  satisfying the usual conditions. It consists of one riskless savings account with interest rate $r>0$ and one risky asset satisfying the following stochastic differential equation (SDE)
\begin{eqnarray*}
&&dS_t=\mu S_tdt+\sigma S_tdW_t,
\end{eqnarray*}
where $\mu>0$ is  the stock growth rate, and $\sigma>0$ is the stock volatility.

Let $(X_t)_{0\leq t\leq T}$  denote the wealth process and $\pi_t$ the amount of wealth an investor holds in risky asset at time $t$. With continuous self-financing strategy,  the wealth process   $(X_t)_{0\leq t\leq T}$ evolves as
\begin{equation*}
dX_t=rX_tdt+\sigma\pi_t(\theta dt+dW_t),
\end{equation*}
where $\theta=\frac{\mu-r}{\sigma}$ is the market price of risk and  $(\pi_t)_{0\leq t\leq T}$ the portfolio process that is ${\cal F}_t$-progressively measurable and satisfies $E[\int^T_0|\pi_{t}|^2dt]<\infty$.

The optimal investment stopping problem is given by
$$\sup_{\pi,\tau}E\left[e^{-\beta\tau}U(X_{\tau}^{0,x,\pi}-K)\right],$$
where  $U$ is a utility function, $\tau\in[0,T]$ is an $\mathscr{F}_t$-adapted stopping time,  $\beta>0$  the utility discount factor,  $K>0$ the minimum wealth threshold value.  If $K=0$ then the problem is a standard utility maximization with investment and stopping. It turns out that  $K>0$ and $K=0$ would lead to completely different optimal trading strategies for non-HARA utility, which indicates that one cannot simply change a portfolio insurance problem into a standard utility maximization by setting $K=0$ to get a seemingly simplified and equivalent problem,  see Examples \ref{exm-nonhara} and  \ref{exam-K=0} for detailed discussions.
\begin{assum}\label{ass-1}
$U\in C^2$ is an increasing and strictly concave function on $[0,\infty)$,  satisfying $U(0)=0$,  $U(\infty)=\infty$, $U'(0)=\infty$,  $U'(\infty)=0$,
$U(x)< C(1+x^p)$  for $x\geq 0$, where $C>0,\, 0<p<1$ are constants,  and  $U(x)=-\infty$ for $x<0$.
\end{assum}

Define the  value function as
\begin{equation*}
V(t,x)=\sup_{\tau,\pi} E\left[e^{-\beta(\tau-t)}U(X_{\tau}^{t,x,\pi}-K)|X_t=x\right]
\end{equation*}
for $(t,x)\in(0,T)\times(K,\infty)$.
Then $V$   satisfies the following HJB variational  equation (see \cite{GUAN2017}):
\begin{eqnarray}\label{main-HJB-1}
&&\min\left\{-{\partial V\over \partial t}-\sup_{\pi}\mathscr{L}^{\pi}[V], V-U(x-K)\right\}=0
\end{eqnarray}
for $(t,x)\in(0,T)\times(K,\infty)$, where
$$\mathscr{L}^{\pi}[V]=rxV_x-\beta V+\pi (\mu-r)V_x+\frac{1}{2}\pi^2 \sigma^2 V_{xx} ,$$
$V_x$ denotes $\frac{\p}{\p x}V(t,x)$, $V$ and  $V_{xx}$, are defined similarly.  The boundary and terminal conditions are given by
\begin{eqnarray} \label{principal problem 3}
 &&V(t,K)=0,\  t \in (0,T),\quad V(T,x)=U(x-K), \ x \in (K,\infty).
\end{eqnarray}
Suppose that $V(t,\cdot)$ is strictly concave,
 then the maximum of $\mathscr{L}^{\pi}[V]$ is achieved at
\begin{equation}\label{controlP}
\pi_t^*=-\frac{\theta}{\sigma}\frac{V_x}{V_{xx}},
\end{equation}
and \eqref{main-HJB-1} is equivalent to
\begin{eqnarray}\label{principal problem 1}
&&\min\left\{-{\partial V\over \partial t}+\frac{\theta^2}{2}\frac{V_x^2}{V_{xx}}-rxV_x+\beta V,\ V-U(x-K)\right\}=0
 \end{eqnarray}
for $(t,x)\in (0,T)\times(K,\infty)$.


We use the dual method to solve the variational equation (\ref{principal problem 1}).
The dual function  of $U(\cdot - K)$ is defined by
\begin{equation*}
\tilde{U}_K(y):=\sup_{x>K}\left[U(x-K)-xy\right]=\tilde U_0(y)-Ky, \quad y>0,
\end{equation*}
where $\tilde U_0$ is the dual function of $U$. It is easy to check that $\tilde{U}_K$ is continuously differentiable, decreasing, strictly convex,   $\tilde{U}_K(0)=\infty$ and
\begin{equation}\label{im-1}
-Ky\leq \tilde{U}_K(y)\leq \tilde{C}+\tilde{C} y^{\frac{p}{p-1}}-Ky,
\end{equation}
where $\tilde{C}=\max\left\{C,(Cp)^{\frac{1}{p-1}}[p^{-1}-1]\right\}$.

Define the dual value function as
\begin{equation*}\label{eq-112}
\tilde{V}(t,y)=\sup_{t\leq\tau\leq T}E\left[e^{-\beta(\tau-t)}\tilde{U}_K(Y_{\tau})|Y_t=y\right],
\end{equation*}
where $(Y_t)_{0\leq t\leq T}$ is a dual process satisfying the SDE
\begin{equation}\label{eq-114}
dY_t=(\beta-r)Y_t dt-\theta Y_td W_t.
\end{equation}
Then the dual   value function satisfies the following variational    equation (see \cite{GUAN2017}):
\begin{equation}\label{dual problem 1}
\min\left\{-{\partial \tilde V\over \partial t}-\frac{\theta^2}{2}y^2\tilde{V}_{yy}-(\beta-r)y\tilde{V}_y+\beta\tilde{V},\ \tilde{V}-\tilde{U}_K\right\}=0
\end{equation}
for $(t,y)\in (0,T)\times(0,\infty)$,
with the terminal condition  given by
$$\tilde{V}(T,y)=\tilde{U}_K(y),\quad y\in (0,\infty).$$
Define
\begin{equation*}
z=\log y, \  \tau=\frac{\theta^2}{2}(T-t),  \  v(\tau,z)=\tilde{V}(t,y).
\end{equation*}
Then $v$ satisfies the following variational equation:
\begin{equation}\label{eq-2}
\min\left\{L[v], \ v-g\right\}=0
\end{equation}
for $(\tau,z)\in \Omega_T:=(0,\theta^2 T/2)\times\mathbb{R}^1$, with the initial condition given by  $v(0,z)=g(z)$ for $ z\in \mathbb{R}^1$,
where
\begin{equation}\label{eq-4}
L[v]=v_{\tau}-v_{zz}+\kappa v_z+\rho v,\ \quad g(z)=\tilde{U}_K(e^z),
\end{equation}
and  constants $\nu, \rho, \kappa$ are defined by
$$ \nu=\frac{2r}{\theta^2}, \ \rho=\frac{2\beta}{\theta^2}, \ \kappa=\nu-\rho+1.$$

The next result shows the existence of a unique  solution of the variational equation  (\ref{eq-2}) with monotonicity properties for each variable. Denote by
$W^{1,2}_{p}(\Omega_T)$ the  Sobolev space and $W^{1,2}_{p, loc}(\Omega_T)$ the local Sobolev space defined by  $W^{1,2}_{p, loc}(\Omega_T):=\{v\in W^{1,2}_p(Q),\ \forall Q\subset\subset\Omega_T\}$.

\begin{thm}\label{Theorem 4.1}
\begin{flushleft}
Problem \eqref{eq-2} has a unique solution $v\in C(\bar{\Omega}_T)\cap W^{1,2}_{p, loc}(\Omega_T)$ for $1<p<+\infty$, satisfying
\begin{equation}\label{eq-106}
g(z)\leq v(\tau,z)\leq \tilde{C}(e^{B\tau+\frac{p}{p-1}z}+1),\ (\tau,z)\in\Omega_T,
\end{equation}
where
$B=|(\frac{p}{p-1})^2-\kappa \frac{p}{p-1}-\rho|+1$ and
$\tilde{C}=\max\left\{C,(Cp)^{\frac{1}{p-1}}[1/p-1]\right\}$.
Furthermore, $v$ satisfies
\begin{equation}\label{derivative-inequality}
v_z\leq 0, \quad - v_z+ v_{zz}> 0,  \quad v_{\tau}\geq 0, \quad  (\tau,z)\in \Omega_T.
\end{equation}
\end{flushleft}
\end{thm}
\proof  See Appendix. \endproof

Since $\tilde V(t,y)=v(\tau,z)$, using Theorem \ref{Theorem 4.1}, we can easily derive the corresponding results for $\tilde V$.

\begin{corollary} \label{cor2.3}
Problem \eqref{dual problem 1}  has a unique solution
$\tilde V\in C([0,T]\times (0,\infty))\cap W^{1,2}_{p, loc}([0,T)\times (0,\infty))$ for $1<p<+\infty$,
satisfying
\begin{equation*}\label{eq-106V}
\tilde U_K(y)\leq \tilde V(t,y)\leq \tilde{C}(e^{B_1(T-t)}y^{{p\over p-1}}+1),\quad (t,y)\in [0,T)\times (0,\infty),
\end{equation*}
where $B_1=B\theta^2/2$ and $B,\, \tilde C$ are given in Theorem \ref{Theorem 4.1}.
Furthermore, $\tilde V$ is decreasing in $t$ and decreasing and strictly convex in $y$, satisfying
\begin{equation}\label{eqn2.20}
\lim_{y\rightarrow 0}-\tilde{V}_y(t,y)=+\infty,\
 \lim_{y\rightarrow \infty}-\tilde{V}_y(t,y):=a\leq K,\ t\in(0,T).
\end{equation}
\end{corollary}
 \proof  See Section \ref{proofs}.
  \endproof

\begin{remark}
We can easily find a strong solution $V$ to the variational HJB equation \eqref{principal problem 1} with conditions \eqref{principal problem 3} by defining
\begin{equation}\label{eq-80}
V(t,x)=\inf_{y>0}[\tilde{V}(t,y)+xy]
\end{equation}
for $t\in(0,T)$ and $x\in(K,\infty)$, and $V$ is strictly increasing and strictly concave in $x$,  see \cite{JIAN2014} for details.
\end{remark}


\section{Main Results}\label{sec:formulas}
In this section, we consider the dual utility function of the form
\begin{equation}\label{main dual utility function}
\tilde{U}_K(y)=\sum_{j=1}^{J}-\frac{1}{q_j}y^{q_j}-Ky,
\end{equation}
where $q_1<q_2<\cdots<q_J<0$.
\begin{exam}\label{ex-utility}
If $J=1$ and $q_1=\frac{\gamma}{\gamma-1}$ with $0<\gamma<1$, then $\tilde{U}_0(y)$ is the dual function of the power utility $U(x)=\frac{1}{\gamma}x^{\gamma}$.
If $J=2$ and $q_1=-3,\ q_2=-1$, then  $\tilde{U}_0(y)$ is the dual function of the non-HARA utility
$$U(x)=\frac{1}{3}H^{-3}(x)+H^{-1}(x)+xH(x),$$
where $H(x)=(\frac{2}{-1+\sqrt{1+4x}})^{1/2}$,  see \cite{BIAN2015}.
\end{exam}
Define $\phi:=L[g]$, where $L$ is defined in (\ref{eq-4}). Direct computation gives
\begin{equation}\label{eqn-10}
\phi(z)=L[g](z)=\sum_{j=1}^JA_je^{q_jz}-\nu Ke^z,
\end{equation}
where $A_j=q_j-\kappa-\rho/q_j$. Note that $A_1<A_2<\cdots<A_J$.

Define the continuation region in $z$-coordinate to be
$\mathcal{C}_z:=\{(\tau,z);\,v(\tau,z)>g(z),\,0<\tau\leq \theta^2 T/2\}$ and the exercise region to be
$\mathcal{S}_z:=\{(\tau,z);\,v(\tau,z)=g(z),\,0<\tau\leq \theta^2 T/2\}$.  We need the following assumption for our main results.
\begin{assum}\label{parameter-assumption}
The parameters of the model satisfy  $K>0$ and
$A_1>0$.
\end{assum}

Now we can prove the existence of the free boundary.
\begin{thm}\label{Theorem-5.2}
Let  Assumption  \ref{parameter-assumption} hold.
Then there exists a unique free boundary $z(\tau)$ defined by
\begin{equation}\label{eq-24}
z(\tau):=\inf\{z;\, v(\tau,z)>g(z)\},\ 0<\tau\leq \theta^2 T/2.
\end{equation}
such that the continuation region $\mathcal{C}_z$ and the  exercise region $\mathcal{S}_z$ can be written respectively as
\begin{equation}\label{eqn-13}
\mathcal{C}_z=\left\{(\tau,z);\, z>z(\tau),\,0<\tau\leq \theta^2 T/2\right\}
\end{equation}
and
\begin{equation}\label{eqn-14}
\mathcal{S}_z=\left\{(\tau,z);\, z\leq z(\tau),\,0<\tau\leq \theta^2 T/2\right\}.
\end{equation}
\end{thm}
\proof    See Section \ref{proofs}.
\endproof

\begin{figure}[!htbp]
\centering
\subfigure[]{
\begin{minipage}{6cm}
\centering
\includegraphics[width=6cm]{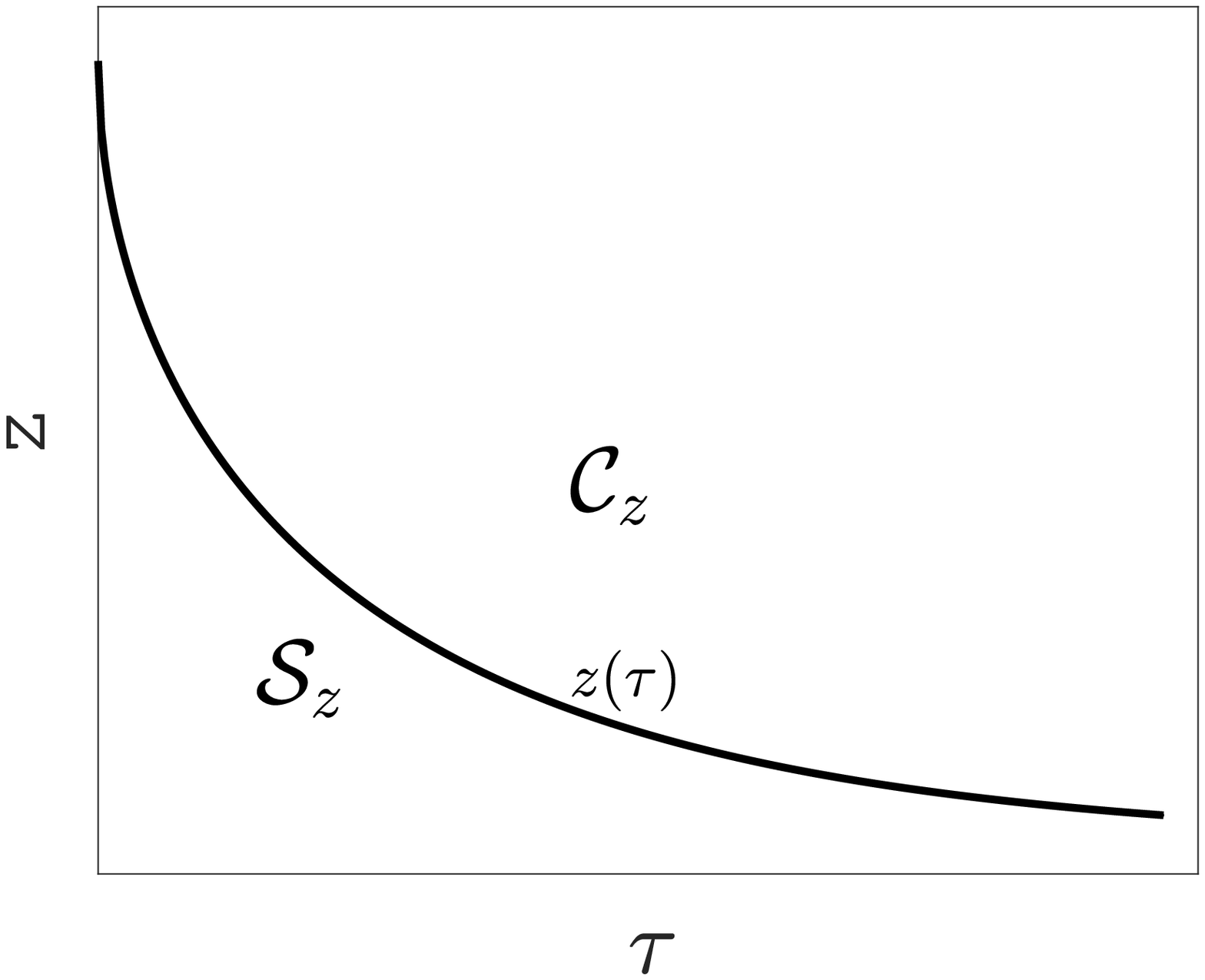}
\end{minipage}
}
\subfigure[]{
\begin{minipage}{6cm}
\centering
\includegraphics[width=6cm]{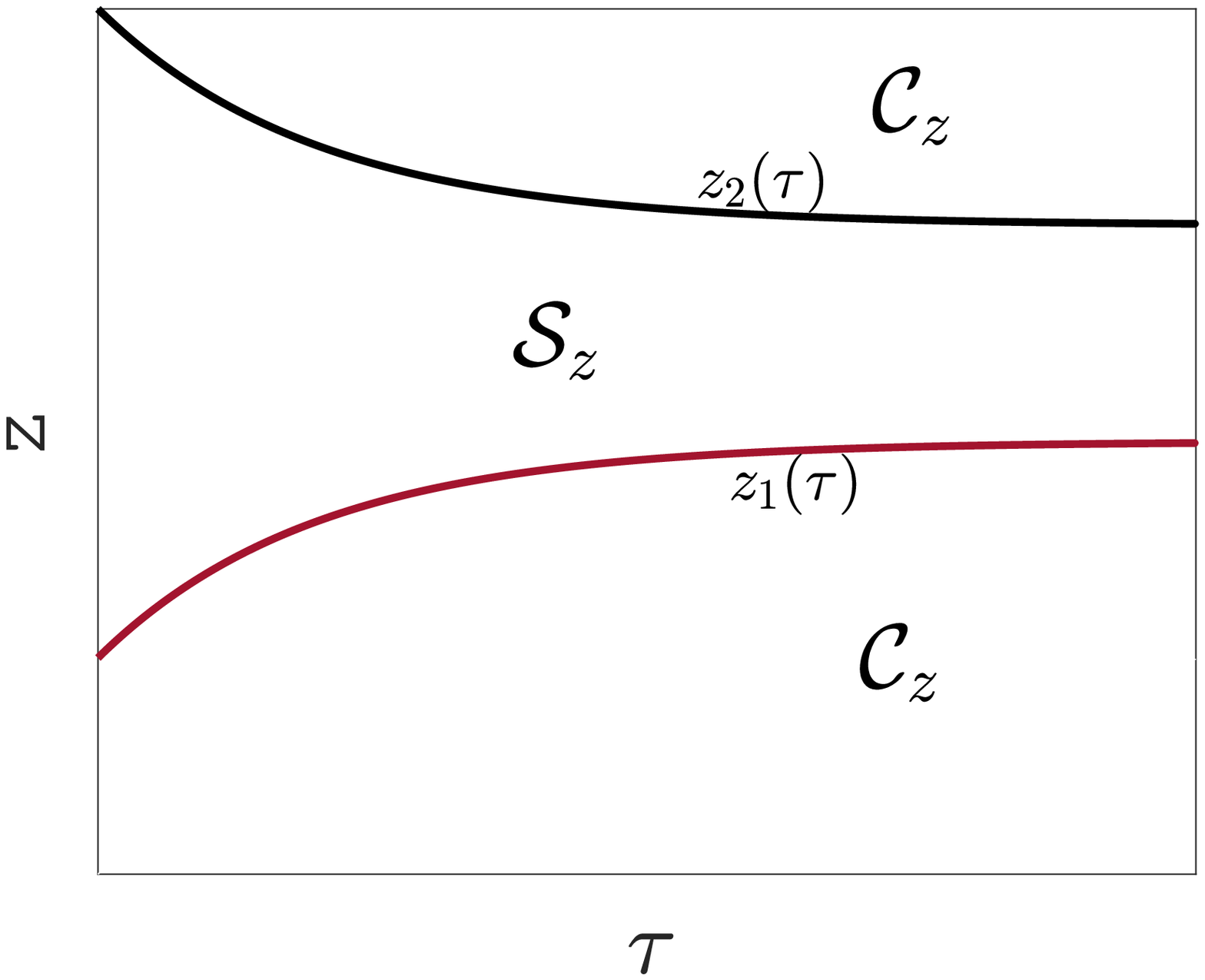}
\end{minipage}
}
\subfigure[]{
\begin{minipage}{6cm}
\centering
\includegraphics[width=6cm]{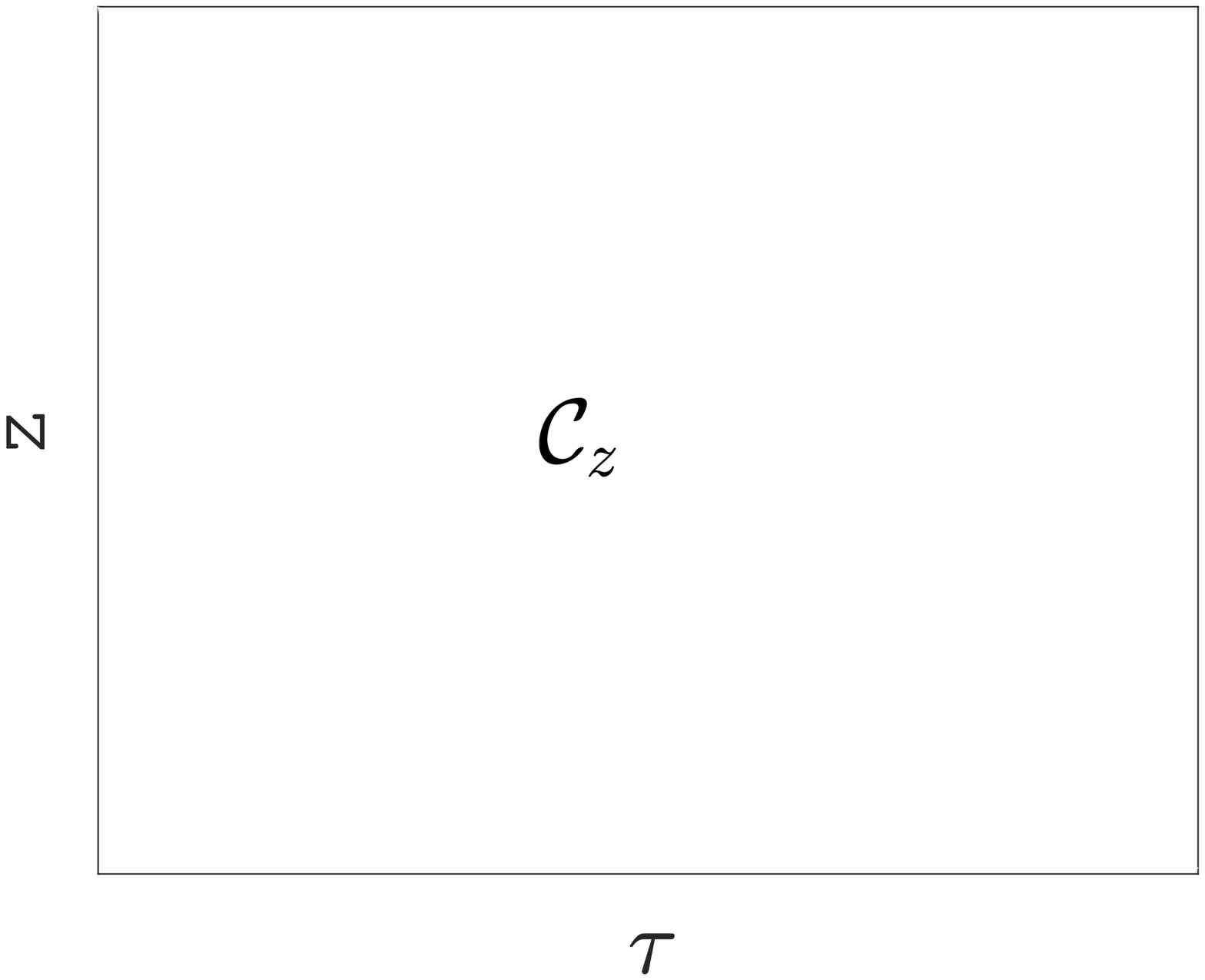}
\end{minipage}
}
\subfigure[]{
\begin{minipage}{6cm}
\centering
\includegraphics[width=6cm]{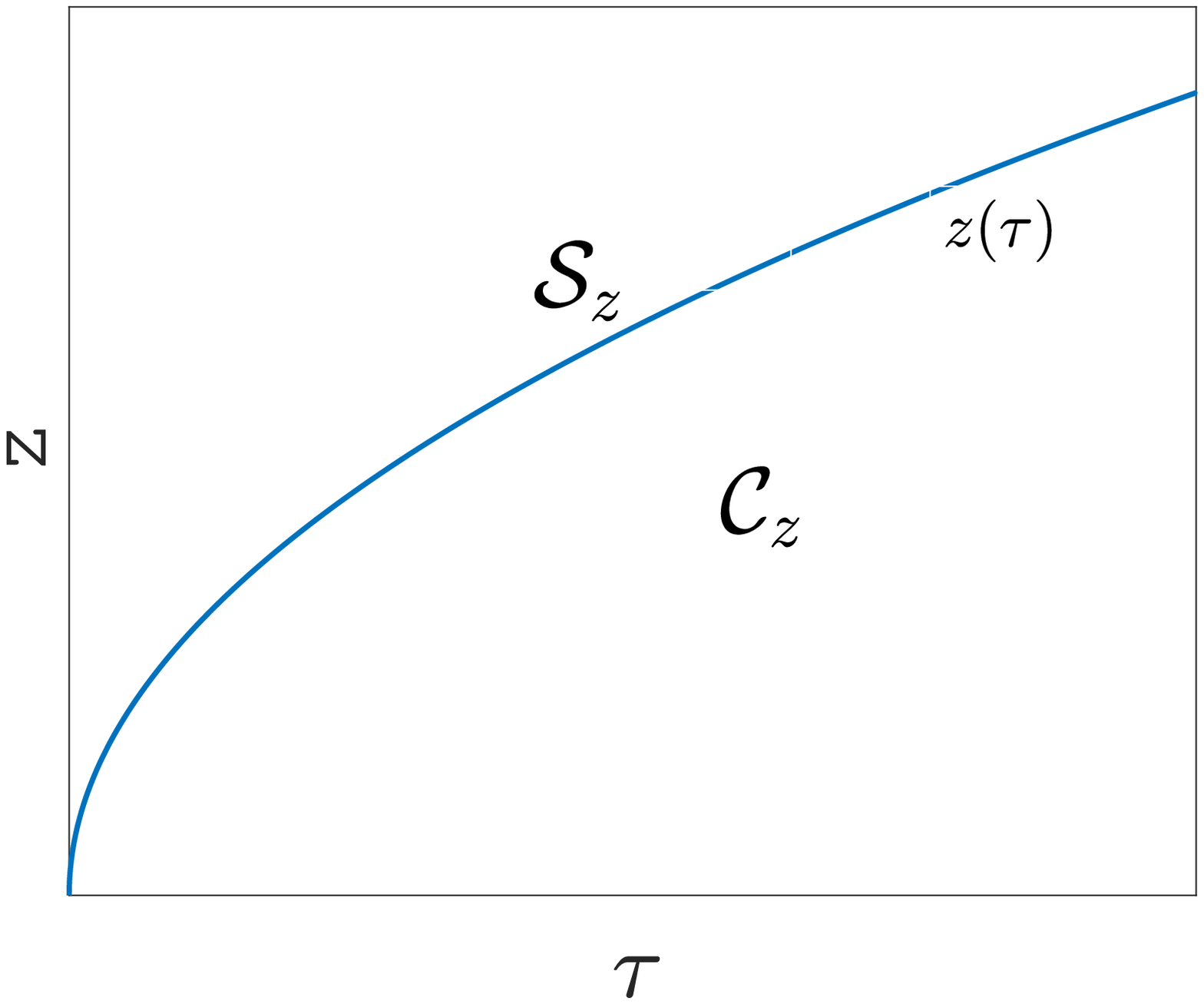}
\end{minipage}
}
\caption{(a) $K>0$, $A_1\geq 0$; \
(b) $K>0$, $A_1<0<A_2,\, A_2^2+4A_1\nu K>0;$ (c) $K>0$, $A_2\leq 0\, \text{or} \, A_1<0<A_2$, $A_2^2+4A_1\nu K\leq 0$;  (d) $K=0$, $A_1<0<A_2$.}
\label{fig-1} \label{example3.3}
\end{figure}

\begin{exam}\label{exm-nonhara}
In this example, we consider  non-HARA utility  ($J=2,\, q_1=-3,\, q_2=-1$ in (\ref{main dual utility function}))  for $K>0$.
Since $A_1<A_2$, we discuss the following three cases.
\begin{itemize}
\item[] Case 1: $A_1\geq 0$. There exists a unique free boundary $z(\tau)$ defined by (\ref{eq-24}).
\item[]Case 2: $A_1< 0<A_2$ and $A_2^2+4A_1\nu K>0$.
There exist two free boundaries $z_1(\tau)$ and $z_2(\tau)$ defined by
\begin{equation}\label{eqt-4}
z_1(\tau):=\inf\{z;\,v(\tau,z)=g(z)\},\ 0<\tau\leq \theta^2 T/2,
\end{equation}
and
\begin{equation}\label{eqt-5}
z_2(\tau):=\sup\{z;\,v(\tau,z)=g(z)\},\ 0<\tau\leq \theta^2 T/2,
\end{equation}
such that the continuation region and the exercise region are given by
\begin{equation}\label{eqt-6}
\mathcal{C}_z=\left\{(\tau,z);\, z<z_1(\tau) \ \hbox{or} \ z>z_2(\tau),\, 0<\tau\leq \theta^2 T/2 \right\}
\end{equation}
and
\begin{equation}\label{eqt-7}
\mathcal{S}_z=\left\{(\tau,z);\, z_1(\tau)\leq z\leq z_2(\tau),\,0<\tau\leq\theta^2 T/2 \right\}.
\end{equation}
Moreover, $z_1(\tau)$ is increasing and $z_2(\tau)$ decreasing with limits
\begin{equation}\label{z_1-0}
\lim_{\tau\to 0}z_1(\tau)=-\frac{1}{2}\log\frac{-A_2-\sqrt{A_2^2+4A_1 K\nu}}{2A_1},
\end{equation}
and
\begin{equation}\label{z_2-0}
\lim_{\tau\to 0}z_2(\tau)=-\frac{1}{2}\log\frac{-A_2+\sqrt{A_2^2+4A_1 K\nu}}{2A_1}.
\end{equation}
\item[]Case 3: $A_2\leq 0$ or $A_1<0<A_2$ and $A_2^2+4A_1\nu K\leq 0$. There is no free boundary and it is not optimal to stop before the maturity.
\end{itemize}
Since the proof is slightly technical, we leave it in Section \ref{proofs}.
Figure \ref{example3.3} (a) - (c) illustrates the three cases discussed above with $\mathcal{C}_z$ the continuation region and $\mathcal{S}_z$ the exercise region.

\end{exam}

\begin{remark}\label{rem-nonhara} For Example \ref{exm-nonhara}, simple algebra shows that $A_i=a_i\beta+b_i$, $i=1,2$, where $a_1=8/(3\theta^2)$, $a_2=4/\theta^2$, $b_1=-2(2+r/\theta^2)$, $b_2=-2(1+r/\theta^2)$. Denote by
$$ \beta_1:=-{b_1\over a_1}=\frac{3}{2}\theta^2+\frac{3}{4}r, \quad
\beta_2:=-{b_2\over a_2}={1\over 2}\theta^2 + {1\over 2}r.$$
Then $A_1\geq 0$ is equivalent to $\beta\geq \beta_1$ and $A_2\leq 0$ is equivalent to $\beta\leq\beta_2$.
For the case $A_1<0<A_2$, or $\beta_2<\beta<\beta_1$, we need to check the sign of $A_2^2+4A_1\nu K$, which requires a more detailed but still simple analysis. Denote by
\begin{eqnarray*}
\beta_3&:=&\beta_2+\sqrt{rK\left(\frac{4}{3}\theta^2+\frac{1}{3}r\right) +\frac{4}{9}r^2K^2}-\frac{2}{3}rK,\\
\beta_4&:=&\beta_2-\sqrt{rK\left(\frac{4}{3}\theta^2+\frac{1}{3}r\right) +\frac{4}{9}r^2K^2}-\frac{2}{3}rK.
\end{eqnarray*}
It is easy to check that $\beta_4<\beta_2<\beta_3<\beta_1$. It turns out that $A_2^2+4A_1\nu K\leq 0$ is equivalent to $\beta_4\leq \beta\leq \beta_3$. Combining the discussions above, we conclude that
the parameter condition of Case 1 in Example \ref{exm-nonhara} is equivalent to $\beta\geq \beta_1$,
that of Case 2  to $\beta_3<\beta<\beta_1$, and that  of Case 3   to $0<\beta\leq \beta_3$.
Recall that  $\beta$ is the utility discount factor. We see that when $\beta$ is small ($\beta\leq \beta_3$), there is no free boundary; when $\beta$ is in the middle ($\beta\in (\beta_3,\beta_1)$), there are two free boundaries; when $\beta$ is large ($\beta\geq \beta_1$), there is one free boundary. The threshold values $\beta_3$ and $\beta_1$ are critical in deciding different optimal trading strategies.

\end{remark}

 The next example is to characterize the optimal exercise and continuation regions for the non-HARA utility discussed in Example \ref{exm-nonhara} when the portfolio insurance value $K$ is set to be 0.
\begin{exam}\label{exam-K=0}
We assume $K=0$ and the same non-HARA utility as in Example \ref{exm-nonhara}. In this case we have $\beta_3=\beta_2$.
\begin{itemize}
\item[] Case 1: $A_1\geq 0$ (equivalently $\beta\geq\beta_1$). There is no free boundary and it is optimal to stop immediately.
\item[] Case 2: $A_1<0<A_2$ (equivalently $\beta_2<\beta<\beta_1$). There exists a unique free boundary defined by
\begin{equation}\label{K=0-def-z}
z(\tau):=\inf\{z; v(\tau,z)=g(z)\}, \ 0<\tau\leq \theta^2 T/2.
\end{equation}
Moreover, $z(\tau)$ is increasing with limits
\begin{eqnarray}
\label{K=0-lim-tau-to-0}
&&\lim_{\tau\to 0} z(\tau)=\frac{1}{2}\log \Big(-\frac{A_1}{A_2}\Big),\\
\label{K=0-lim-tau-infty}
&&\lim_{\tau\to\infty}z(\tau)=\infty.
\end{eqnarray}
\item[] Case 3: $A_2\leq 0$ (equivalently $\beta\leq\beta_2$). There is no free boundary and    it is not optimal to stop before the maturity.
\end{itemize}
We leave the proof in Section \ref{proofs}. Figure \ref{example3.3} (d) illustrates the  Case 2 discussed above with $\mathcal{C}_z$ the continuation region and $\mathcal{S}_z$ the exercise region.
\end{exam}

Examples \ref{exm-nonhara} and \ref{exam-K=0} show that there is a fundamental difference on optimal trading strategies with  $K=0$ and $K>0$. For example, when $A_1<0<A_2$, there exists a unique free boundary for $K=0$ whereas there exist either two free boundaries or no free boundary for $K>0$, which  implies that one has to use different optimal trading strategies in the presence of portfolio insurance $K>0$ and cannot simply set $K=0$ to reduce the problem into a standard utility maximization problem.

With  Assumption \ref{parameter-assumption}, we can directly verify that $\phi(z)$ defined by \eqref{eqn-10} is strictly decreasing and there exists a unique $z_0\in\mathbb{R}$ such that
\begin{equation}\label{eqn-12}
\phi(z_0)=0.
\end{equation}
\begin{thm}\label{Theorem 3.1}
Let Assumption \ref{parameter-assumption} hold. Then the free boundary
$z(\tau)$ defined by  \eqref{eq-24}
 is strictly decreasing with ${\displaystyle\lim_{\tau\rightarrow 0}}z(\tau)=z_0$,
where $z_0$ is defined by \eqref{eqn-12},
 and $z(\tau)\in C[0,\theta^2 T/2]\cap C^{\infty}(0,\theta^2 T/2]$.
 Furthermore, $z(\tau)$ satisfies the following integral equation
\begin{equation}\label{integral equation 1}
-\int^{\infty}_{z_0}G(\tau,z(\tau)-y)\phi(y)dy+\int^{\tau}_0G(\tau-s,z(\tau)-z(s)) \phi(z(s))z'(s)ds=0,
\end{equation}
where  $G$ is  the Green function defined by
\begin{equation} \label{green}
G(\tau,z)=\frac{1}{\sqrt{4\pi\tau}}\exp\left(-\frac{(z-\kappa \tau)^2}{4\tau}-\rho\tau\right).
\end{equation}
\end{thm}
\proof See Section \ref{proofs}. \endproof

In the following, we conduct the asymptotic analysis of the free boundary and construct the global approximation for the dual problem.
We investigate the asymptotic behaviour of the free boundary near the expiry by using the integral equation (\ref{integral equation 1}).
\begin{thm}\label{Theorem 3.2}
Let  Assumption \ref{parameter-assumption} hold. Then the free boundary $z(\tau)$ defined by \eqref{eq-24}, for $0<\tau<<1$, satisfies approximately
\begin{equation*}\label{eq-41}
z(\tau)\approx z_0-2A\sqrt{\tau},
\end{equation*}
where $A$ is a positive solution of the following equation
\begin{equation}\label{eq-56-transedental}
\frac{1}{2}e^{-A^2}-\frac{\sqrt{\pi}}{2}A+A^2\int_0^1e^{-A^2\eta^2}\frac{3\eta^2+\eta^4}{(1+\eta^2)^2}d\eta=0.
\end{equation}
The numerical solution of equation \eqref{eq-56-transedental} is  $A\approx 0.56292056798247$.
\end{thm}
\proof See Section \ref{proofs}.
\endproof

The next result gives the asymptotic property of the free boundary $z(\tau)$ defined by \eqref{eq-24} as time to maturity $\tau$ tends to infinite.
\begin{thm}\label{Theorem 4.4}
Let Assumption \ref{parameter-assumption} hold and $z^*$ be the unique solution of the equation
$$\sum_{j=1}^J-\frac{1}{q_j}(q_j-\lambda)e^{(q_j-1)z}-K(1-\lambda)=0,$$
where $\lambda=\frac{1}{2}(\kappa -\sqrt{\kappa^2+4\rho})$. Then  the free boundary defined by \eqref{eq-24} satisfies
$$\lim_{\tau\rightarrow\infty}z(\tau)=z^*.$$
\end{thm}
\proof See Section \ref{proofs}. \endproof

\begin{exam}
Simple calculation shows that, for power utility  ($J=1$ in (\ref{main dual utility function})), we have
\begin{equation*}\label{eqn-16}
z_0=\frac{1}{q_1-1}\log\frac{K\nu}{A_1},\quad z^*=\frac{1}{q_1-1}\log \frac{Kq_1(1-\lambda)}{\lambda-q_1},
\end{equation*}
and, for non-HARA utility  ($J=2,\,q_1=-3,\, q_2=-1$ in (\ref{main dual utility function})), we have
\begin{equation*}\begin{split}\label{eqn-17}
z_0&=-\frac{1}{2}\log\frac{-A_2+\sqrt{A_2^2+4A_1\nu K}}{2A_1},\\
z^*&=-\frac{1}{2}\log\frac{-(q_2-\lambda)+\sqrt{(q_2-\lambda)^2+\frac{4}{3} (q_1-\lambda)(1-\lambda)}}{\frac{2}{3}(q_1-\lambda)}.
\end{split}\end{equation*}
By Assumption \ref{parameter-assumption} and $q_j-\lambda>0,\, j=1,2$ (see \eqref{q-lambda}), we
can  verify that the expressions inside the above logarithmic functions are positive.

\end{exam}

Now we seek a simple  approximation formula for $z(\tau)$ defined by \eqref{eq-24} such that (i) it has asymptotic expansion $z_0-2A\sqrt{\tau}$ for small  $\tau$ and (ii) it
approaches $z^*$ for large $\tau$.  For this, we seek an approximation of the form
$$z_*(\tau):=z_0-2A\sqrt{\frac{1-e^{-b\tau}}{b}},$$
where $b>0$.  To make it match with the large $\tau$ behaviour, we need $b=\frac{4A^2}{(z_0-z^*)^2}$.   Hence, the global closed-form approximation of the free boundary $z(\tau)$ defined by \eqref{eq-24}  is given by
\begin{equation}\label{global approximation}
z_*(\tau):= z_0-(z_0-z^*)\sqrt{1-e^{-b^*\tau}}, \ \ b^*=\frac{4A^2}{(z_0-z^*)^2}.
\end{equation}

The next result presents a global closed-form approximation of the free boundary for  problem \eqref{principal problem 1} with conditions \eqref{principal problem 3}.

\begin{thm}\label{Theorem 5.1}
Let the dual utility function be given by \eqref{main dual utility function} and Assumption \ref{parameter-assumption} hold. Let
 $z_*(\tau)$ in \eqref{global approximation} be the global closed form approximation (GCA) to the free boundary $z(\tau)$ defined by \eqref{eq-24}.  Then the unique free boundary of problem \eqref{principal problem 1} with condition \eqref{principal problem 3} is strictly decreasing and approximately determined by
\begin{equation*}
x(t)= -\tilde{U}_K'\left(\exp\left(z_*(\theta^2(T-t)/2)\right)\right),\ 0\leq t\leq T.
\end{equation*}
Furthermore,  the primal value function is given by
$$ V(t,x)=\tilde{V}(t,I(t,x))+xI(t,x),$$
and  the optimal feedback control is given by
\begin{equation}\label{controlD}
  \pi^*_t=\frac{\theta}{\sigma}I(t,x)\tilde{V}_{yy}(t,I(t,x)),
  \end{equation}
where $\tilde V$ is the dual value function, approximately given by
\begin{equation} \label{approxDV}
 \tilde{V}(t,y) =\tilde{U}_K(y)-\int_0^{\tau} \int_{ z_*(s)}^{\infty}G\left(\tau-s,\ln y-w\right)\phi(w)dwds,
\end{equation}
 $\tau=\theta^2(T-t)/2$, and $y=I(t,x)$ is the unique solution of the equation $\tilde V_y(t,y)+x=0$ for $x>K$.

\end{thm}
 \proof See Section \ref{proofs}. \endproof

\begin{remark} In the continuation region, the optimal feedback control $\pi^*$ can be computed  either with (\ref{controlP}) using the primal value function or with (\ref{controlD}) using the dual value function.
The two methods would produce the same optimal trading strategy due to the strong duality relation. It is in general more difficult to find the primal value function than to find the dual value function as the former satisfies a nonlinear PDE in the continuation region whereas the latter a linear PDE. The dual value function has an integral representation which makes possible computing the optimal control, provided the dual free boundary is known. This is where  the GCA plays a pivotal role. It would be virtually impossible without the GCA to determine the optimal control in the continuation region as both the primal and dual value functions depend on unknown free boundaries.
\end{remark}

\section{Numerical Examples}\label{sec:examples}

In this section, we compare the  numerical results derived using the global closed-form approximation (GCA) and the binomial tree method (BTM). We now briefly explain to use BTM to solve our problem. BTM cannot be directly applied to solve the original investment stopping problem, however, it can be used to solve the dual optimal stopping problem which is essentially an American options pricing problem with one additional difficulty, that is, one has to find the initial value $y$ of the dual process from the equation $\tilde V_y(t,y)+x=0$ while $\tilde V$ is to be determined.  To circumvent the problem, we use the following procedure.

First, we fix an arbitrary $y_0>0$ and build a binomial tree for the dual process $Y$ up to time $T$ and then use the dynamic programming method to solve the dual optimal stopping problem and find the value
$\tilde V_y(0,y_0)$  at time 0. We then check the sign of  $\tilde V_y(0,y_0)+x$: if positive, we decrease the value of  $\tilde V_y(0,y_0)$ by setting  $y_1=y_0/10$; if  negative,   we increase the value of  $\tilde V_y(0,y_0)$ by setting  $y_1=10y_0$. Repeat the process and get $\tilde V_y(0,y_1)$. If $\tilde V_y(0,y_1)+x$ and $\tilde V_y(0,y_0)+x$ have the same sign,  we  set $y_0=y_1$ and repeat the process above; if they have different signs,  we have found an interval, bounded by $y_0$ and $y_1$, that contains   a solution to the equation  $V_y(t,y)+x=0$. We then use the bisection method to find $y$ with linear  convergence.
Once the initial value $y$ for the dual process is determined, we can get easily the value $\tilde V(0,y)$ and the free boundary for the dual problem. Finally, using the dual relation, we can find the optimal value and the free boundary for the primal problem.

\begin{exam}\label{ex-6.1}
 We discuss the free boundary  and the optimal strategy of the optimal investment stopping problem \eqref{principal problem 1} with conditions \eqref{principal problem 3}  for power utility and non-HARA utility defined   in Example \ref{ex-utility}.

The parameters used are  $\mu=0.1$, $\beta=0.1$, $r=0.05$, $\sigma=0.3$, $K=1$, $\gamma=0.5$, $T=1$. The number of time steps for binomial tree method is $N=700$, which gives $4$ decimal point accuracy. These parameters satisfy Assumption \ref{parameter-assumption}.

  In Figure \ref{fig-1} we plot the optimal exercise boundaries for power and non-HARA utilities using both the global closed-form approximation (GCA) and the binomial tree method (BTM). It is clear that the GCA and the BTM produce the free boundaries with the same shape and very small gaps.
In Figure  \ref{fig-2} we depict the sample paths of the optimal wealth and the corresponding optimal trading strategy using the GCA for power and non-HARA utilities. We can see that the optimal trading strategy becomes zero after time $\tau_0$, the first time the optimal wealth process hits the free boundary before the terminal time $T$, and $\tau_0$ is the optimal stopping time of investing in risky assets and  the optimal wealth becomes $X_t=X_{\tau_0}e^{r(t-\tau_0)}$ for $\tau_0\leq t\leq T$.

\begin{figure}[!htbp]
\centering
\subfigure[]{
\begin{minipage}{6.5cm}
\centering
\includegraphics[width=6.5cm]{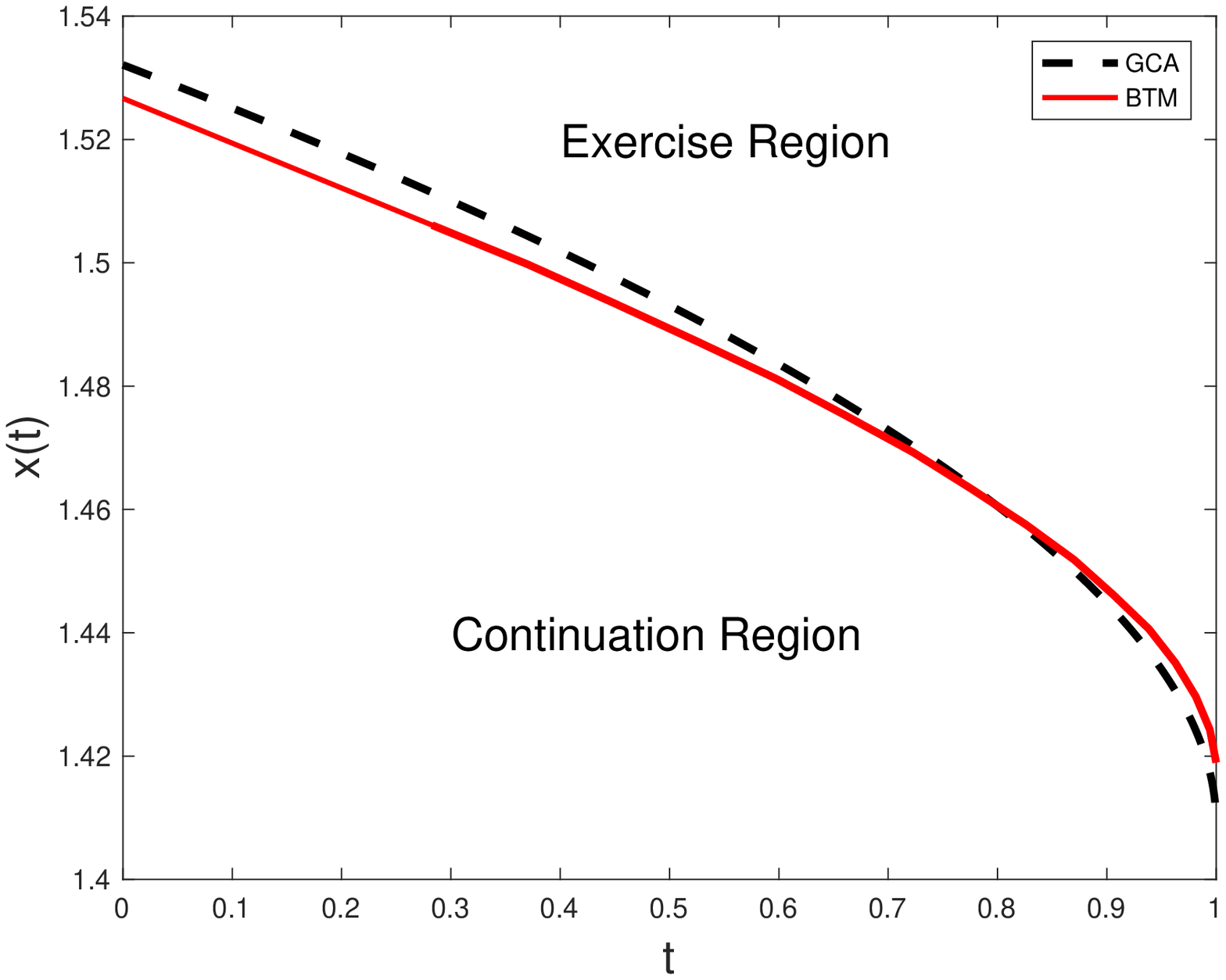}
\end{minipage}
}
\subfigure[]{
\begin{minipage}{6.5cm}
\centering
\includegraphics[width=6.5cm]{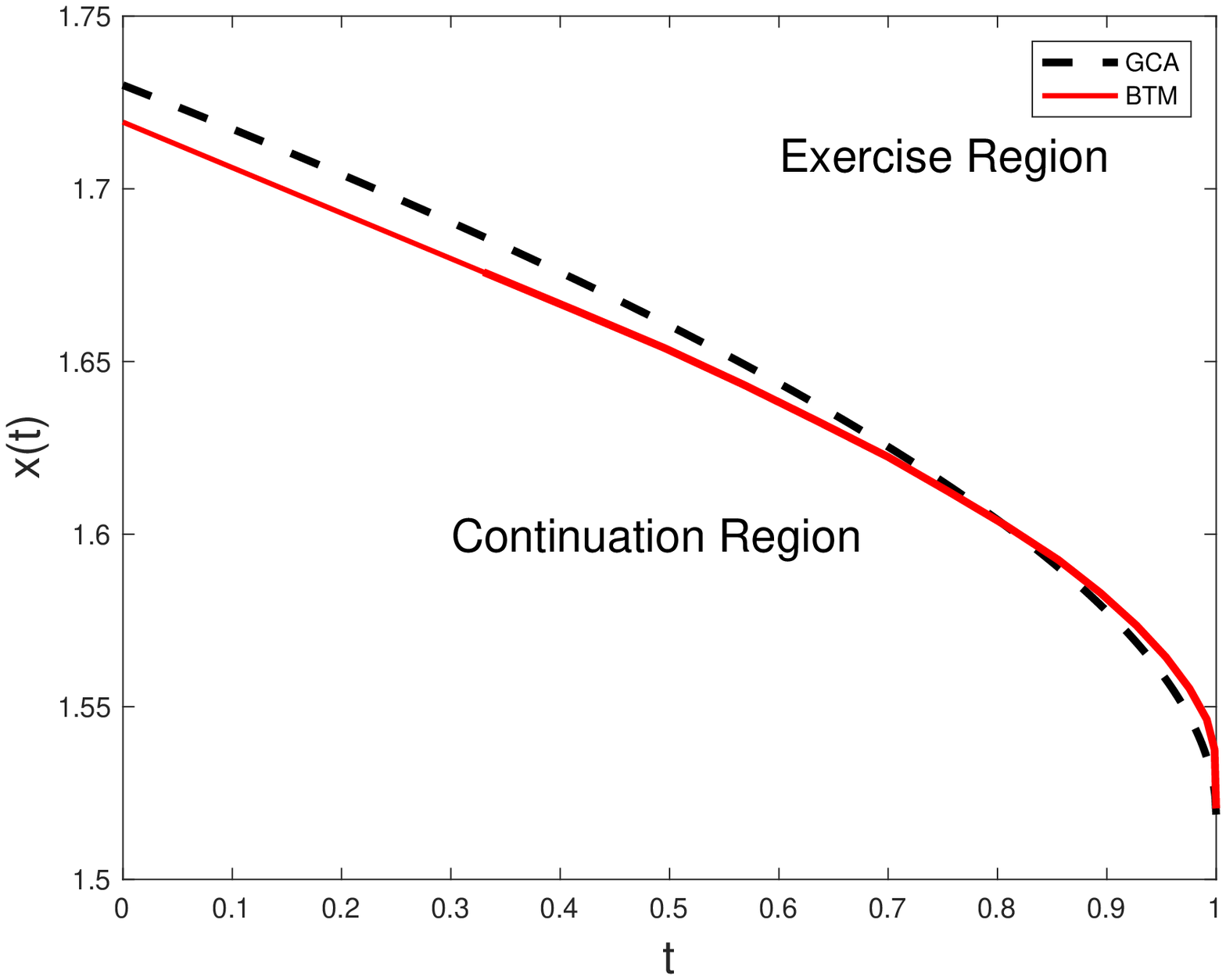}
\end{minipage}
}
\caption{(a) The optimal exercise boundary compared with BTM for power utility; \
(b) The optimal exercise boundary compared with BTM for non-HARA utility. }
\label{fig-1}
\end{figure}

\begin{figure}
\centering
\subfigure[]{
\begin{minipage}[b]{6.5cm}
\includegraphics[width=6.5cm]{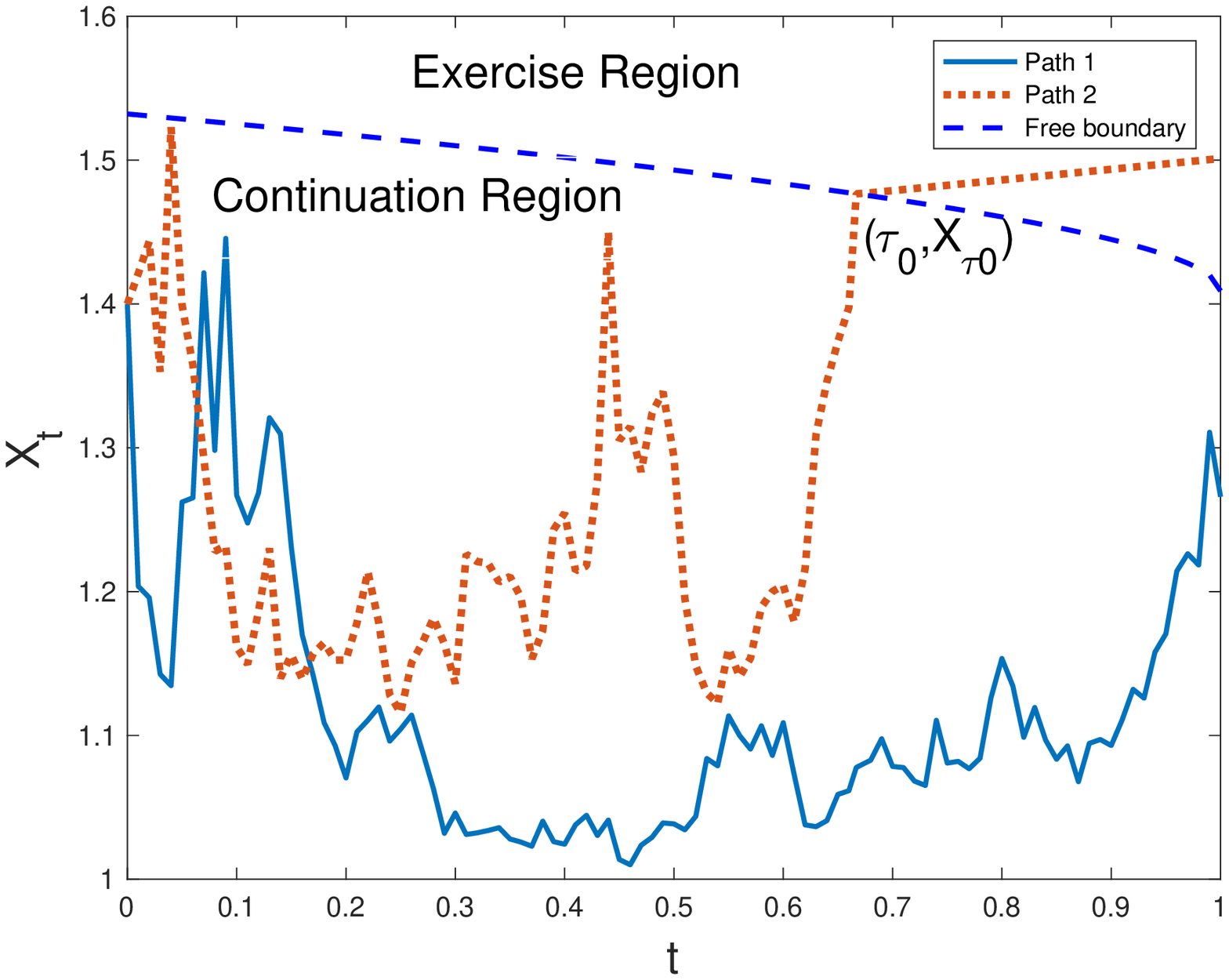}\\
\includegraphics[width=6.5cm]{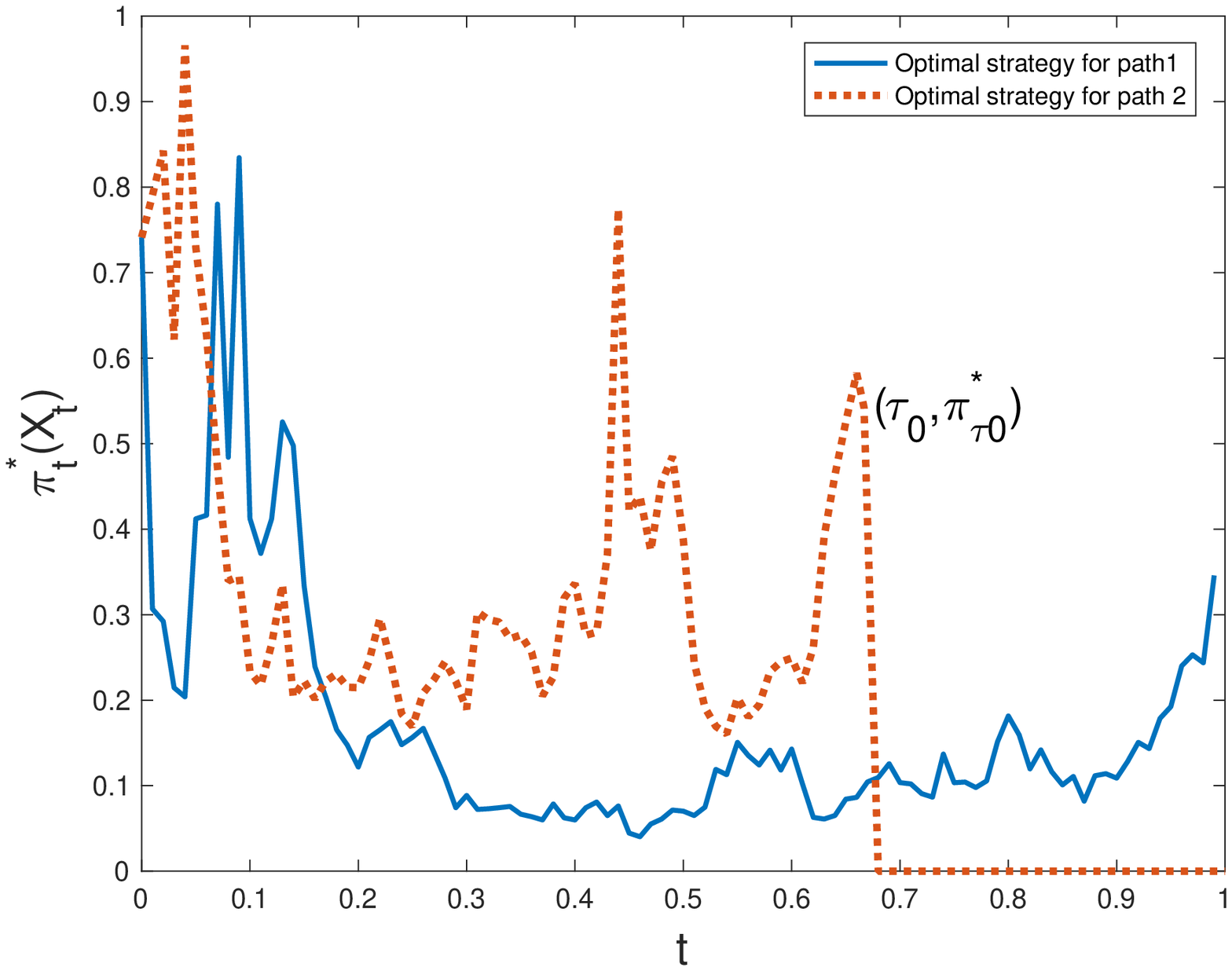}
\end{minipage}
}
\subfigure[]{
\begin{minipage}[b]{6.5cm}
\includegraphics[width=6.5cm]{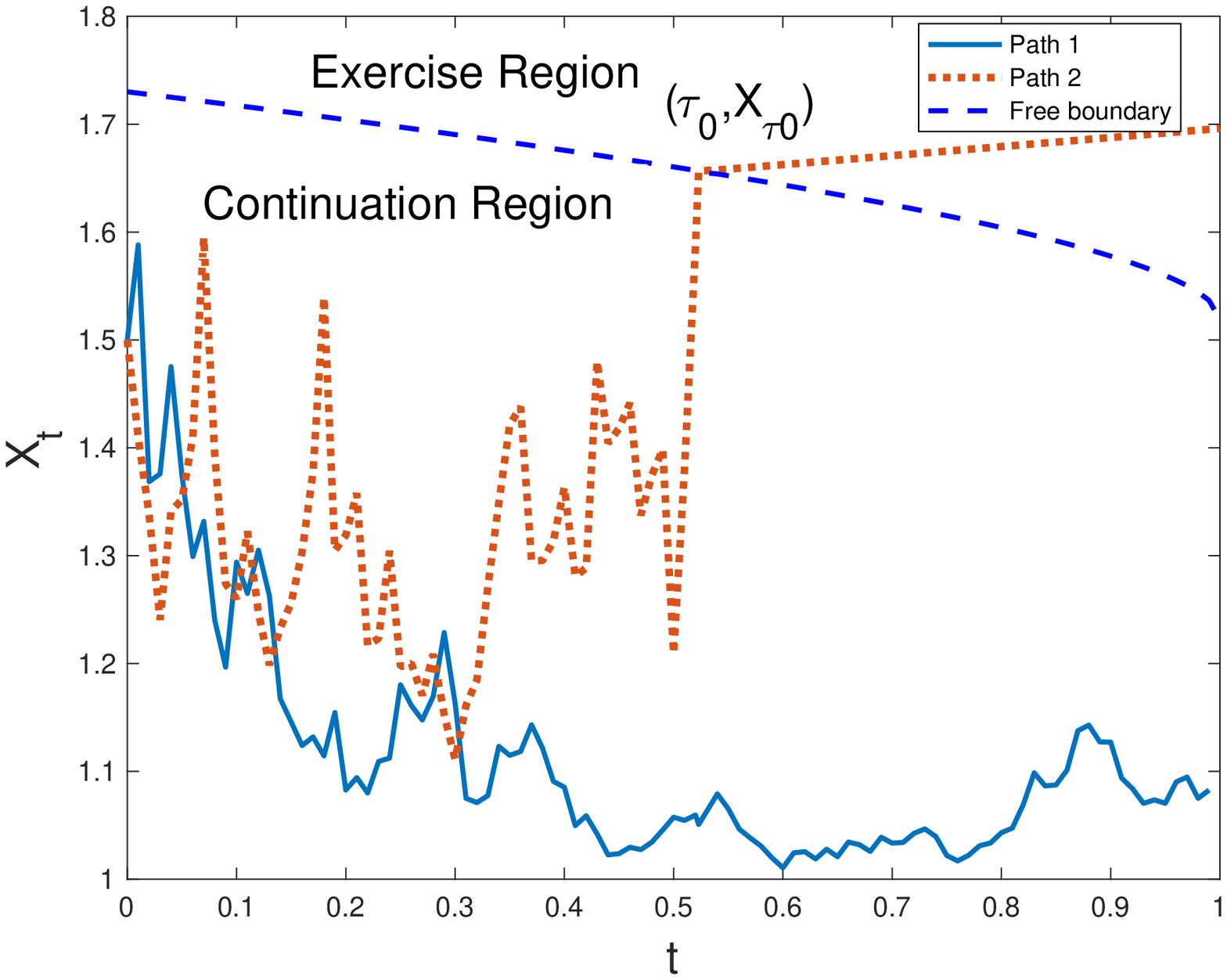}\\
\includegraphics[width=6.5cm]{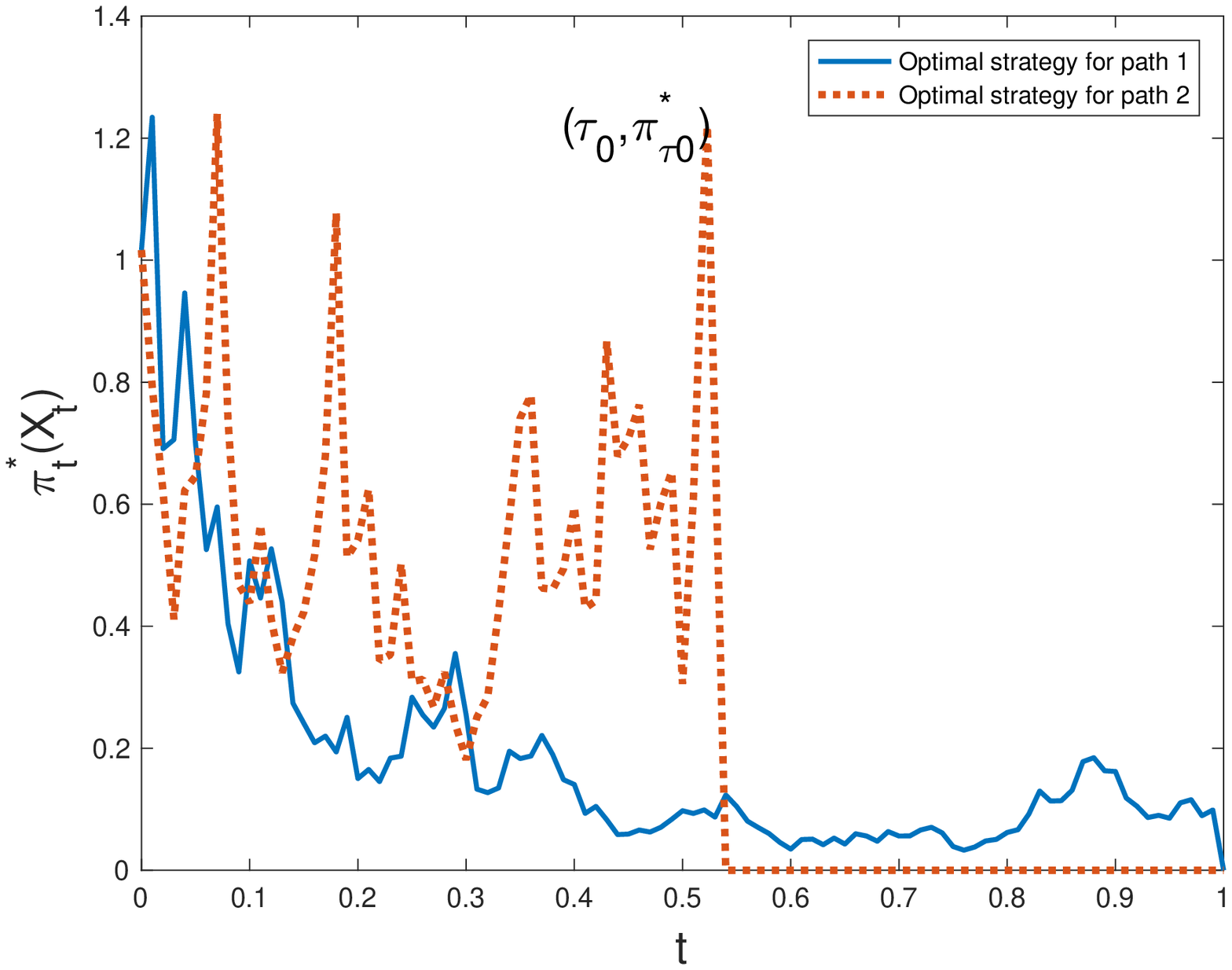}
\end{minipage}
}
\caption{ (a) Two different sample paths of wealth with initial wealth $x_0=1.4$  and optimal strategy for power utility.
 (b) Two different sample paths  of wealth with initial wealth $x_0=1.5$ and optimal strategy for non-HARA utility.
}
\label{fig-2}
\end{figure}

\end{exam}

\begin{exam}\label{ex-6.2}
In this example, we compare the optimal values and the optimal strategies obtained by the closed-form approximation and the binomial tree method at the initial time.
\begin{itemize}
\item[(i)]
For power and non-HARA utility, we compare the numerical results between GCA and BTM.
The parameters used are $\mu=0.1$, $\beta=0.1$, $r=0.05$, $\sigma=0.3$, $K=1$, $\gamma=0.5$, $T=1$, $t=0,$ initial wealth $x_0=1.5$, number of time steps for binomial tree approach $N=700$. The numerical result is shown in Table \ref{table-1}.

\item[(ii)]
In Table \ref{table-2}, we give the mean and standard deviation of the absolute and relative difference between BTM and GCA for power
and non-HARA utility. We fix $K=1$, $T=1$, $t=0,$ initial wealth $x_0=1.5$, and number of time steps for binomial tree approach $N=700$. The rest parameters are selected randomly: $10$ samples of $\mu$ from the uniform distribution on interval $[0.05,0.15]$, $r$
 on $[0.02,0.08]$, $\beta$ on $[0.05,0.15]$, $\sigma$ on $[0.10,0.40]$, $\gamma$ on $[0.2,0.6]$.   We also require the parameters satisfy Assumption \ref{parameter-assumption}.

\end{itemize}

From the numerics in Table \ref{table-2}, we observe that the difference between the GCA and BTM optimal values is very small,  whereas the computational time for GCA is much less than that for BTM. The GCA is shown to be correct and fast.  Compared to the optimal values, the error for computing the optimal strategies using both the BTM and GCA is bigger. This is not surprising, as the optimal strategies are computed with the derivatives of the value functions.

\begin{table}[!htbp]
\caption{Comparison between GCA and BTM for Example \ref{ex-6.2} (i).} 
\centering 
\begin{tabular}{l c c c c}\label{table-1}\\
\hline
 \multirow{2}{*}{}   &\multicolumn{2}{c}{Power utility} & \multicolumn{2}{c}{Non-HARA utility}\\
  \cline{2-5}
 & Optimal value & Optimal strategy &   Optimal value & Optimal strategy \\
\hline
GCA value&1.4128&0.6558&1.5094&0.6776\\
BTM value&1.4031&0.7454&1.5116&0.6846\\
\hline
 Difference&0.0096 & 0.0899 &0.0022& 0.0069\\
Relative difference&0.0069& 0.1206&0.0015& 0.0101\\
\hline
 Time for GCA& 22.7s &10.9s &11.4s & 5.6s\\
 Time for BTM&1683.2s&1664.2s&2777.6s&2744.6s\\
\hline
\end{tabular}
\end{table}

\begin{table}[!htbp]
\caption{
Comparison between GCA and BTM for Example \ref{ex-6.2} (ii).
 }
\centering 
\begin{tabular}{l c c c c}\label{table-2}\\
\hline
 \multirow{2}{*}{}   &\multicolumn{2}{c}{Power utility} & \multicolumn{2}{c}{Non-HARA utility}\\
  \cline{2-5}
 & Optimal value & Optimal strategy &   Optimal value & Optimal strategy \\
\hline
 Avg difference&0.0074&0.0969&0.0034&0.0281\\
Std difference&0.0050&0.1495&0.0062&0.0462\\
Avg relative difference&0.0050&0.0745&0.0022&0.0630\\
Std relative difference&0.0033&0.1145&0.0045&0.0919\\
\hline
Avg time for GCA&23.2s&8.2s&23.0s&9.3s\\
Avg time for BTM&2640.9s&2609.6s&2878.6s&2844.2s\\
\hline
\end{tabular}
\end{table}

\end{exam}

\section{Proofs } \label{proofs}
In this section we give detailed proofs of the results of the paper.

 \subsection{Proof of Corollary \ref{cor2.3}}
 \proof  Everything is a straightforward translation of Theorem \ref{Theorem 4.1}. We only need to show
(\ref{eqn2.20}) holds. For some fixed $y_0>0$ and $y<y_0$,
using  $\tilde U_K(0)=\infty$ and the convexity of $\tilde V$ in $y$, we have
$$\tilde{V}_y(t,y)\leq\frac{\tilde{V}(t,y_0)-\tilde{V}(t,y)}{y_0-y} \leq\frac{\tilde{V}(t,y_0)-\tilde{U}_K(y)}{y_0-y},$$
which gives ${\displaystyle \lim_{y\rightarrow 0}}\left(-\tilde{V}_y(t,y)\right)=+\infty$.
Similarly, for some fixed $y_0>0$ and $y_0<y$, using (\ref{im-1}), we obtain
\begin{eqnarray*}
&&0\leq-\tilde{V}_y(t,y)\leq-\frac{\tilde{V}(t,y)-\tilde{V}(t,y_0)}{y-y_0} \leq-\frac{\tilde{U}_K(y)-\tilde{V}(t,y_0)}{y-y_0}\leq-\frac{-Ky-\tilde{V}(t,y_0)}{y-y_0},
\end{eqnarray*}
which gives ${\displaystyle \lim_{y\rightarrow \infty}}\left(-\tilde{V}_y(t,y)\right):=a\leq K$.
  \endproof

 \subsection{Proof of Theorem  \ref{Theorem-5.2}}
\proof
In the exercise region, we immediately have
$v_z= g_z$.
In the continuation region,  since
$v_z(0,z)=g_z$ and $v_z(\tau,z)=g_z$ for $(\tau,z)\in\p\mathcal{C}_z$,
and by Assumption  \ref{parameter-assumption} and $q_{j}<0$, we have
\begin{equation*}\label{eq-111}
L[g_z]=\phi'(z)=\sum_{j=1}^JA_j q_je^{q_jz}-K\nu e^z\leq 0.
\end{equation*}
 On the other hand, in the continuation region it holds that $L[v_{z}]=0$. So we have $L[v_{z}-g_{z}]\geq 0$ in the continuation region. By comparison, we obtain that
\begin{equation*}\label{eqn-15}
v_z-g_z\geq 0.
\end{equation*}
As a consequence, if $(\tau,z_1)\in \mathcal{C}_z$,  i.e., $v(\tau,z_1)>g(z_1)$,  then for any $z_2>z_1$,
$$v(\tau,z_2)-g(z_2)\geq v(\tau,z_1)-g(z_1)>0,$$
from which we infer that $(\tau,z_2)\in \mathcal{C}_z$. This indicates each $\tau$-section of $\mathcal{C}_z$ is connected. The existence of the free boundary $z(\tau)$ now follows. We obtain (\ref{eq-24}) and (\ref{eqn-13}). Moreover,
(\ref{eqn-14}) follows from (\ref{eqn-13}).
\endproof

\subsection{Proof of Example \ref{exm-nonhara}}
\proof
 Case 1:
If $A_1>0$,  from Theorem \ref{Theorem-5.2}, we know  there exists a unique free boundary $z(\tau)$ defined by (\ref{eq-24}).
If $A_1=0,$ then  $A_2>A_1=0,$ Theorem \ref{Theorem-5.2} implies that   there exists a unique free boundary $z(\tau)$ defined by (\ref{eq-24}).

 Case 2: We now prove \eqref{eqt-4} - \eqref{eqt-7}. Denote that
$\Lambda:=\{(\tau,z);\, z_1(\tau)\leq z\leq z_2(\tau),\,0<\tau\leq\theta^2 T/2 \}$.
Since $A_1<0<A_2,$ $A_2^2+4A_1\nu K>0$ and $\phi(z)=e^z(A_1 e^{-4z}+A_2 e^{-2z}-\nu K)$, then there exists two roots for equation $\phi(z)=0$. We denote the two roots  by $z_{I},\ z_{II}$ with $z_{I}<z_{II}$.
By a direct computation, we have
$$z_I=-\frac{1}{2}\log\frac{-A_2-\sqrt{A_2^2+4A_1 K\nu}}{2A_1},$$
$$z_{II}=-\frac{1}{2}\log\frac{-A_2+\sqrt{A_2^2+4A_1 K\nu}}{2A_1}.$$
Then from the definition of the exercise region $\mathcal{S}_z=\{(\tau,z);\,v(\tau,z)=g(z),\,0<\tau\leq\theta^2 T/2\}$ and the variational equation \eqref{eq-2}  , we have $L[v]=L[g]=\phi(z)\geq 0$ for $(\tau,z)\in \mathcal{S}_z.$   This implies that
$$
\mathcal{S}_z\subseteq\left\{(\tau,z);\,\phi(z)\geq 0,\,0<\tau\leq\theta^2 T/2\right\}=\left\{0<\tau\leq\theta^2 T/2,\ z_{I}\leq z\leq z_{II}\right\}.
$$
This shows that the $\tau$-section $\{z;\,v(\tau,z)=g(z),\,0<\tau\leq\theta^2 T/2\}$ of the exercise region $\mathcal{S}_z$ is bounded. Therefore, $z_1(\tau)$ and $z_2(\tau)$ in \eqref{eqt-4} - \eqref{eqt-5} are well defined. By the definitions of $z_1(\tau)$ and $z_2(\tau)$, we obtain that $\mathcal{S}_z\subseteq\Lambda$.
Now, we prove that
\begin{equation}\label{eqt-8}
\Lambda \subseteq \mathcal{S}_z.
\end{equation}
Since
$$
\left\{ (\tau,z);\,  z=z_1(\tau)\, or \, z=z_2(\tau),\, 0<\tau\leq\theta^2 T/2\right\}\subseteq \mathcal{S}_z\subseteq \left\{(\tau,z);\,\phi(z)\geq 0,\,0<\tau\leq\theta^2 T/2\right\},
$$
we have
$\Lambda\subseteq \left\{(\tau,z);\,\phi(z)\geq 0,\,0<\tau\leq\theta^2 T/2\right\}$.
Assume that \eqref{eqt-8} is false. Then there exists a non-empty subset  $\mathscr{N}=\mathcal{C}_z\cap\Lambda$ and the parabolic boundary $\p_p\mathscr{N}\subseteq \bar{\Omega}_T-\mathcal{C}_z$. Here $\bar{\Omega}_T$ denotes the closure of $\Omega_T$. Thus
\begin{eqnarray*}
&&L[v]=0, \ (\tau,z)\in\mathscr{N},\\
&&
L[g]=\phi(z)\geq 0, \ (\tau,z)\in\mathscr{N},\\
&&
v=g, \ (\tau,z)\in\p_p\mathscr{N}.
\end{eqnarray*}
By the comparison principle, $v\leq g$ in $\mathscr{N}$, which implies that $\mathscr{N}=\emptyset$.  Hence the contradiction arises.
Therefore, \eqref{eqt-8} holds. So $\Lambda=\mathcal{S}_z$, i.e., \eqref{eqt-7} holds true.
 \eqref{eqt-6} follows  from \eqref{eqt-7}.

Next, we prove the monotonicity of the two free boundaries. If $z_1(\tau)$ is not increasing, there exist $\tau_1<\tau_2$ such that $z_1(\tau_1)> z_1(\tau_2)$. Since $v_{\tau}\geq 0$ (see \eqref{derivative-inequality}), we have
$$g(z_1(\tau_2))=v(\tau_2, z_1(\tau_2))\geq v(\tau_1,z_1(\tau_2))>g(z_1(\tau_2)),$$
which is a contradiction.
Hence, $z_1(\tau)$ is increasing. Similarly, $z_2(\tau)$ is decreasing.

Finally, we prove \eqref{z_1-0} and \eqref{z_2-0}. If
$$\lim_{\tau\to 0}z_1(\tau)>z_{I},$$
then for any $z$ satisfying
$$\lim_{\tau\to 0}z_1(\tau)>z>z_{I},$$
and $\tau=0$,
we have
\begin{eqnarray*}
0&=&v_{\tau}-v_{zz}+\kappa v_z+\rho v\\
  &=&v_{\tau}+\phi(z)>0,
\end{eqnarray*}
where the last inequality follows from the fact that $v_{\tau}\geq 0$ and $\phi(z)>0$ for
$z_I<z<z_{II}$.
This is a contradiction. Hence, \eqref{z_1-0} holds. By a similar argument, we can obtain
\eqref{z_2-0}.

Case 3:
In fact,  if $A_2\leq 0$ or $A_1<0<A_2$, $A_2^2+4A_1\nu K\leq 0$,  then
$L[g]=\phi(z)=e^z(A_1 e^{-4z}+A_2 e^{-2z}-\nu K)\leq 0$.
Hence, $g$ is a subsolution of problem
\begin{eqnarray}\label{spp-3}
&&L[v]=0, \ (\tau,z)\in \Omega_T,\\
\label{spp-4}
&&v(0,z)=g(z),\ z\in \mathbb{R}^1.
\end{eqnarray}
Denote the solution of the problem (\ref{spp-3}) - (\ref{spp-4}) by $\tilde{v}$. Then by comparison, we obtain that
$$\tilde{v}-g\geq 0\ \text{in}\ \Omega_{T}.$$
Therefore, $\tilde{v}$ is also the solution of problem (\ref{eq-2}).
\endproof

\subsection{Proof of Example \ref{exam-K=0}}
\proof
Case 1 and Case 3 can be easily proved as follows: Since $K=0$, we have $L[g]=\phi(z)=A_1 e^{-3z}+A_2 e^{-z} \geq 0$ if $A_1\geq 0$ or $\phi(z)\leq 0$ if $A_2\leq 0$ due to the relation $A_1<A_2$.  If $\phi(z)\leq 0$, then by the same argument as in the proof of Example  \ref{exm-nonhara}, we conclude that there is no free boundary and it is not optimal to stop before the maturity. If $\phi(z)\geq 0$, then $v=g$  is the solution to problem  \eqref{eq-2}, which implies that there is no free boundary and it is optimal to stop immediately.

Next we prove Case 2. We can show \eqref{K=0-def-z}, \eqref{K=0-lim-tau-to-0} and monotonicity of $z(\tau)$ following a similar argument as  in the proof of   Example  \ref{exm-nonhara}.  We only  need to prove  \eqref{K=0-lim-tau-infty}. If $z(\tau)$ is bounded, then we have $\displaystyle{\lim_{\tau\to\infty}} z(\tau)<\infty$.  Denote $\displaystyle{\lim_{\tau\to\infty}} z(\tau):= a$.

We rewrite problem \eqref{eq-2} as
\begin{eqnarray*}
&&L[v]=I_{\{z\geq z(\tau)\}}\phi(z),\ (\tau,z)\in\Omega_T,\\
&&v(0,z)=g(z),\ z\in\mathbb{R},
\end{eqnarray*}
where  $I_A$ is the indicator function of set $A$.
By Green's identity, we have
\begin{equation*}\label{value-function-integral-1}
v(\tau,z)=\int_{-\infty}^{\infty}G(\tau,z-y)g(y)dy+\int_0^{\tau}\int^{\infty}_{z(\tau-s)} G(s,z-y)\phi(y)dyds,
\end{equation*}
where $G$ is the Green function defined by (\ref{green}).
We set
$$\Lambda_1(\tau)=\frac{1}{3} e^{-3z(\tau)-3A_1\tau}+e^{-z(\tau)-A_2\tau},$$
$$\Lambda_2(\tau)=\frac{A_1}{\sqrt{\pi}}e^{-3 z(\tau)}\int_0^{\tau}e^{-3 A_1 s}\int_{\frac{z(\tau-s)-z(\tau)+(\kappa+6)s}{2\sqrt{s}}}^{\infty} e^{-\eta^2}d\eta ds,$$
$$\Lambda_3(\tau)=\frac{A_2}{\sqrt{\pi}}e^{-z(\tau)}\int_0^{\tau}e^{-A_2 s}\int_{\frac{z(\tau-s)-z(\tau)+(\kappa+2)s}{2\sqrt{s}}}^{\infty} e^{-\eta^2}d\eta ds.$$
Since $v(\tau,z(\tau))=g(z(\tau))$, we have
$$\Lambda_1(\tau)+\Lambda_2(\tau)+\Lambda_3(\tau)=g(z(\tau)).$$
By dominated convergence theorem, we have
\begin{eqnarray*}
\lim_{\tau\to\infty}\Lambda_2(\tau)&=& \frac{A_1}{\sqrt{\pi}}e^{-3 a}\int_0^{\infty}e^{-3 A_1 s}\int_{\frac{(\kappa+6)\sqrt{s}}{2}}^{\infty} e^{-\eta^2}d\eta ds,\\
                                                      &=&-\frac{\kappa+6}{6\sqrt{\pi}}e^{-3 a}\int_0^{\infty} e^{-(\frac{\kappa^2}{4}+\rho)t^2} dt <\infty.
\end{eqnarray*}
Since $A_1<0<A_2$, we have
\begin{eqnarray*}
&&\lim_{\tau\to\infty}\Lambda_3(\tau)\leq A_2 e^{- a}\int_0^{\infty}e^{-A_2 s} ds <\infty, \\
&&\lim_{\tau\to\infty}\Lambda_1(\tau)=\infty.
\end{eqnarray*}
As $\displaystyle{\lim_{\tau\to\infty}}g(z(\tau))=g(a)<\infty$,  this leads to a contradiction. Hence, we obtain that $z(\tau)$ is increasing and unbounded, i.e., \eqref{K=0-lim-tau-infty} holds.
\endproof

\subsection{Proof of Theorem \ref{Theorem 3.1}}
\proof

From Theorem $\ref{Theorem-5.2}$, the variational problem \eqref{eq-2}  can be written as
\begin{eqnarray}\label{eq-25}
&&L[v]=0 \ \text{for} \ z>z(\tau),\ 0<\tau\leq \theta^2 T/2, \\
\label{eq-26}
&&v(\tau,z)=g(z) \ \text{for}\ z\leq z(\tau),\ 0<\tau\leq \theta^2 T/2,\\
\label{eq-27}
&&v_z(\tau,z(\tau))=g_z(z(\tau)), \ 0<\tau\leq \theta^2 T/2,\\
\label{eq-28}
&&v(0,z)=g(z), \ z\in\mathbb{R}, \nonumber
\end{eqnarray}
where $L$ and $g$ is defined as in \eqref{eq-4}.

Firstly, we claim that $z(\tau)$ is non-increasing.
Otherwise, there exists some $0<\tau_0<\tau_1$ such that $z(\tau_0)<z(\tau_1)$.  Then since  $v_{\tau}\geq 0$ (see \eqref{derivative-inequality}), we obtain that
$$0=v(\tau_1,z(\tau_1))-g(z(\tau_1))\geq v(\tau_0,z(\tau_1))-g(z(\tau_1))>0, $$
where the second inequality follows from the definition of the free boundary $z(\tau)$.  This leads to contradiction.
Then we claim that
\begin{equation}\label{eq-29}
z(\tau)<z_0\ \text{for} \ \tau>0.
\end{equation}
Let $\bar{v}=v-g$.  We rewrite the variational problem \eqref{eq-2} as
\begin{eqnarray}\label{eq-30}
&&\min\{L[\bar{v}]+\phi(z),\ \bar{v}\}=0,\ (\tau,z)\in\Omega_{T}, \\
\label{eq-31}
&&\bar{v}(0,z)=0,\ z\in\mathbb{R},
\end{eqnarray}
where $\phi(z)$ is defined by (\ref{eqn-10}).

Let $U$  be the solution to
\begin{eqnarray}\label{eq-32}
&&L[U]=-\phi(z),\ (\tau,z)\in\Omega:=\{(\tau,z)\in\Omega_{T};\, z>z_0\}, \\
\label{eq-33}
&&U(\tau,z)=0, \ (\tau,z)\in\p_p\Omega,
\end{eqnarray}
where $\p_p\Omega$ is the parabolic boundary of $\Omega$.

Since $\phi(z)$ is strictly decreasing and $z>z_0$, by \eqref{eqn-12}, we have $L[U]=-\phi(z)>0$ in $\Omega$. By the maximum principle (see \cite[Theorem 2.7]{LIEBERMAN1996}),  we have $U>0$ in $\Omega$. Then  Hopf's lemma (see \cite[Lemma 2.8]{LIEBERMAN1996}) leads to $U_z(\tau,z_0)>0$ for $\tau>0$.
 By (\ref{eq-30}) - (\ref{eq-33}), we have $L[\bar{v}]\geq -\phi(z)= L[U]$ and $\bar{v}(\tau,z)\geq 0 = U(\tau,z)$ for $(\tau,z)\in\p_p\Omega$. By the comparison principle \cite[Corollary 2.5]{LIEBERMAN1996},
 we see that $\bar{v}\geq U>0$ in $\Omega$. This implies that $z(\tau)\leq z_0$. Otherwise, there exists some $z_1\in (z_0,z(\tau))$  such that $\bar{v}(\tau,z_1)=0$.

If there exists some $\tau_0>0$ such that $z(\tau_0)=z_0$, then we have
$$\bar{v}(\tau_0,z(\tau_0))=U(\tau_0,z(\tau_0))=0.$$
Hopf's lemma  (see \cite[Lemma 2.8]{LIEBERMAN1996}) implies that
$$\bar{v}_z(\tau_0,z(\tau_0))>U_z(\tau_0,z(\tau_0))>0.$$
Since $\bar{v}_z(\tau,z(\tau))=0$,  for any $\tau>0$, this leads to contradiction. So (\ref{eq-29}) is proved.

Hence, we have  ${\displaystyle\lim_{\tau\rightarrow 0}}z(\tau)\leq z_0$.
If ${\displaystyle \lim_{\tau\rightarrow 0}}z(\tau)< z_0$,
then for some $z\in({\displaystyle\lim_{\tau\rightarrow 0}z(\tau),z_0)}$, by \eqref{eq-25}, we have
$$L[v]|_{\tau=0}=[v_{\tau}-v_{zz}+\kappa v_z+\rho v]|_{\tau=0}=0.$$
This leads to
$$v_{\tau}|_{\tau=0}=[v_{zz}-\kappa v_z-\rho v]|_{\tau=0}=-\phi(z)<0,$$
where the last inequality follows from $\phi$ is strictly decreasing, $\phi(z_0)=0$ and $z<z_0$.  This contradicts with the fact that $v_{\tau}(0,z)\geq0 $ (see \eqref{derivative-inequality}) 

We now prove that $z(\tau)\in C[0,\theta^2 T/2]$. If this is not true, then there exists some $\tau_0\in[0,\theta^2 T/2)$ such that
$$z_1<z_2, \ \text{where}\ z_1=\lim_{\tau\rightarrow\tau_0^{+}}z(\tau),\ z_2=\lim_{\tau\rightarrow\tau_0^{-}}z(\tau). $$
By (\ref{eq-29}), we have $z_1,\ z_2\leq z_0$.  For any $z\in[z_1,z_2]$,
by \eqref{eq-25}, we have
$$L[v]|_{\tau=\tau_0}=[v_{\tau}-v_{zz}+\kappa v_z+\rho v]|_{\tau=\tau_0}=0.$$
For any $z\in[z_1,z_2]$,  this leads to
$$v_{\tau}|_{\tau=\tau_0}=[v_{zz}-\kappa v_z-\rho v]|_{\tau=\tau_0}=-\phi(z)\leq 0,$$
where the last inequality follows from $\phi$ is decreasing, $\phi(z_0)=0$ and $z_1,\, z_2\leq z_0$.
By \eqref{derivative-inequality}, this means $\phi(z)=0$ for any $z\in[z_1,z_2]$, while $\phi$ is a strictly decreasing function. The contradiction arises. Therefore $z(\tau)\in C[0,\theta^2 T/2]$ is true.
Furthermore we can use the bootstrap argument developed by \cite{FRIEDMAN1975} to conclude that $z(\tau)\in C^{\infty}(0,\theta^2 T/2]$.

To prove the rest of the results in this theorem, by (\ref{eq-26}), we have
\begin{eqnarray}\label{eq-34}
&&v(\tau,z(\tau))=g(z(\tau)), \ \tau>0.
\end{eqnarray}
Differentiating (\ref{eq-34}) in $\tau$, by (\ref{eq-27}), we obtain
\begin{equation}\label{eq-123}
v_{\tau}(\tau,z(\tau))=0.
\end{equation}
Furthermore,  (\ref{eq-25}) implies
$$L[v](\tau,z(\tau))=v_{\tau}(\tau,z(\tau))-v_{zz}(\tau,z(\tau))+\kappa v_z(\tau,z(\tau))+\rho v(\tau,z(\tau))=0,$$
which leads to
\begin{equation}\begin{split}\label{eq-35}
v_{zz}(\tau,z(\tau))&=\kappa g_{z}(z(\tau))+\rho g(z(\tau)),\ \tau>0.
\end{split}\end{equation}
By \eqref{eq-123} and Theorem \ref{Theorem 4.1}, we derive
\begin{eqnarray*}
&&L[v_{\tau}]=0\  \text{in continuation region},\\
&&v_{\tau}(\tau,z(\tau))=0,  \ 0<\tau<\theta^2 T/2,\\
&&v_{\tau}(0,z)\geq 0, \ z\in\mathbb{R}.
\end{eqnarray*}
Hopf's lemma and the maximum principle imply that $v_{\tau}>0$ in the continuation region and $v_{\tau z}>0$ at $(\tau,z(\tau))$.
Differentiating (\ref{eq-27}) in $\tau$, we have
$$v_{\tau z}(\tau,z(\tau))+v_{zz}(\tau,z(\tau))z'(\tau)=g_{zz}(z(\tau))z'(\tau).$$
By \eqref{eq-35},
\begin{equation}\label{eq-92}
v_{z\tau}(\tau,z(\tau))=-\phi(z(\tau))z'(\tau)>0,
\end{equation}
Since $\phi$ is strictly decreasing, by (\ref{eq-29}), we derive that
$$-\phi(z(\tau))<-\phi(z_0)=0,\ \tau>0.$$
Therefore, we obtain $z'(\tau)<0$.

The standard method for finding an integral equation of the free boundary starts with the Green function
$G$ defined by (\ref{green}), which satisfies
$$L[G]=G_{\tau}-G_{zz}+\kappa G_z+\rho G=0.$$
Denote by $u(\tau,z)=v_{\tau}(\tau,z)$ and
\begin{equation}\label{eq-37}
I(z,\tau,s)=\int^{\infty}_{z(s)}G(\tau-s,z-y)u(s,y)dy.
\end{equation}
Note that  ${\displaystyle \lim_{s\rightarrow \tau}}G(\tau-s,z-y)=\delta(z-y)$, where $\delta$ is a Dirac delta function, therefore for any $z>z(\tau)$,
\begin{equation*}\label{eq-38}
\lim_{s\rightarrow \tau}I(z,\tau,s)=u(\tau,z).
\end{equation*}
With this in mind, we can relate the solution $u(\tau,z)$ to the initial condition by integrating $I_s(z,\tau,s)$ between $s=0$ and $s=\tau$.

Differentiating (\ref{eq-37}), also noting $u(s,z(s))=v_s(s,z(s))=0$ (see \eqref{eq-123}),  yields
\begin{equation*}\label{dev-integral-eq}
I_s(z,\tau,s)=-\int^{\infty}_{z(s)}G_{\tau}(\tau-s,z-y)u(s,y)dy+\int^{\infty}_{z(s)}G(\tau-s,z-y)u_s(s,y)dy.
\end{equation*}
Simple computation, using integration by parts, gives
\begin{eqnarray*}
&&\int^{\infty}_{z(s)}G(\tau-s,z-y)u_s(s,y)dy\\
&=&\int_{z(s)}^{\infty} G(\tau-s,z-y)\left[u_{zz}-\kappa u_z-\rho u\right](s,y)dy\\
&=&-G(\tau-s,z-z(s))u_z(s,z(s))+\, \int_{z(s)}^{\infty}[G_{zz}-\kappa G_z-\rho G](\tau-s,z-y)u(s,y)dy\\
&=&-G(\tau-s,z-z(s))u_z(s,z(s))+\int_{z(s)}^{\infty}G_{\tau}(\tau-s,z-y)u(s,y)dy.
\end{eqnarray*}
Hence,
$$I_s(z,\tau,s)=-G(\tau-s,z-z(s))u_z(s,z(s)).$$
Integrating $I_s(z,\tau,s)$ from $s=0$ to $s=\tau$, we obtain
\begin{equation}\label{eq-78}
u(\tau,z)-\int^{\infty}_{z_0}G(\tau,z-y)u(0,y)dy=-\int^{\tau}_0G(\tau-s,z-z(s))u_z(s,z(s))ds.
\end{equation}
Now we calculate $u(0,y)$ for $y\geq z_0$. By (\ref{eq-25}), we have
$$u(0,y)=v_{\tau}(0,y)=v_{zz}(0,y)-\kappa v_z(0,y)-\rho v(0,y)=-L[g]=-\phi(y).$$
Also $u_z(s,z(s))=v_{\tau z}(s,z(s))$ is given by (\ref{eq-92}).
Since $u(\tau,z(\tau))=v_{\tau}(\tau,z(\tau))=0$ (see \eqref{eq-123}), letting $z=z(\tau)$ in \eqref{eq-78}, we have (\ref{integral equation 1}).
\endproof

\subsection{Proof of Theorem \ref{Theorem 3.2}}
\proof
We postulate that
\begin{equation}\label{eq-122-approxi}
z(\tau)=z_0-2A\sqrt{\tau}+o(\sqrt{\tau}),\ \tau\rightarrow 0. \end{equation}
A direct computation shows that the first term in \eqref{integral equation 1} is given by
\begin{eqnarray}\label{eq-43}
-\int^{\infty}_{z_0}G(\tau,z(\tau)-y)\phi(y)dy
&=&\sum_{j=1}^J-\frac{1}{2}A_j
e^{q_jA_j\tau+q_jz(\tau)}\text{erfc}\left(\frac{z_0-z(\tau) +(\kappa -2q_j)\tau}{2\sqrt{\tau}}\right)\nonumber\\
&&{}+\frac{1}{2}\nu Ke^{-\nu\tau+z(\tau)} \text{erfc}\left(\frac{z_0-z(\tau)+(\nu-\rho-1)\tau}{2\sqrt{\tau}}\right),
\end{eqnarray}
where $\text{erfc}(z)$ is the complementary error function defined by
$$
\text{erfc}(z)=\frac{2}{\sqrt{\pi}}\int^{\infty}_z e^{-\eta^2}d\eta.
$$
By Taylor's expansion and \eqref{eq-122-approxi}, we have
\begin{eqnarray}\label{eq-44}
e^{q_jA_j\tau+q_jz(\tau)}&=&
e^{o(\sqrt{\tau})+q_j(z_0-2A\sqrt{\tau}+o(\sqrt{\tau}))}\nonumber\\
&=&e^{q_jz_0}(1-2q_jA\sqrt{\tau}+o(\sqrt{\tau}))\\
e^{-\nu\tau+z(\tau)}&=&e^{z_0}(1-2A\sqrt{\tau}+o(\sqrt{\tau})).\nonumber
\end{eqnarray}
Similarly, Taylor's expansion gives
\begin{eqnarray}\label{eq-46}
\text{erfc}\left(A(1+o(1))+\frac{\kappa -2q_j}{2}\sqrt{\tau}\right)=\text{erfc}\left(A(1+o(1))\right)-\frac{\kappa -2q_j} {\sqrt{\pi}}e^{-A^2}\sqrt{\tau}+o(\sqrt{\tau})
\end{eqnarray}
and
\begin{eqnarray}\label{eq-47}
\text{erfc}(A(1+o(1))+\frac{\nu-\rho-1}{2}\sqrt{\tau})=\text{erfc}(A(1+o(1)))-\frac{\nu-\rho-1}{\sqrt{\pi}}e^{-A^2}\sqrt{\tau}+o(\sqrt{\tau}).
\end{eqnarray}
Since $\phi(z_0)=0$, by \eqref{eq-43} - \eqref{eq-47}, we derive that
\begin{eqnarray}\label{eq-48}
&&-\int^{\infty}_{z_0}G(\tau,z(\tau)-y)\phi(y)dy\nonumber\\
&=&\sum_{j=1}^J-\frac{1}{2}A_je^{q_jz_0}\left[1-2q_jA\sqrt{\tau}+o(\sqrt{\tau})\right]\left[\text{erfc}(A(1+o(1))) -\frac{\kappa -2q_j}{\sqrt{\pi}}e^{-A^2}\sqrt{\tau}+o(\sqrt{\tau})\right]\nonumber\\
&&{}+\frac{1}{2}\nu K e^{z_0}\left[1-2A\sqrt{\tau}+o(\sqrt{\tau})\right]\left[\text{erfc}(A(1+o(1))) -\frac{\nu-\rho-1}{\sqrt{\pi}}e^{-A^2}\sqrt{\tau} +o(\sqrt{\tau})\right]\nonumber\\
&=&-\frac{1}{2}\phi(z_0)\text{erfc}(A(1+o(1))) -\sum_{j=1}^J\frac{1}{2}A_je^{q_jz_0} \left[(-2q_jA\sqrt{\tau})\text{erfc}(A(1+o(1)))-\frac{\kappa -2q_j} {\sqrt{\pi}}e^{-A^2}\sqrt{\tau}\right]\nonumber\\
&&{}+\frac{1}{2}\nu K e^{z_0}\left[(-2A\sqrt{\tau})\text{erfc}(A(1+o(1))) -\frac{\nu-\rho-1}{\sqrt{\pi}}e^{-A^2}\sqrt{\tau}\right] +o(\sqrt{\tau})\nonumber\\
&=&\left(A\,\text{erfc}(A(1+o(1)))-\frac{1}{\sqrt{\pi}}e^{-A^2}\right)\sqrt{\tau}\left[\sum_{j=1}^Jq_jA_je^{q_jz_0}-\nu K e^{z_0}\right]+o(\sqrt{\tau}).
\end{eqnarray}
We use the transformation $t=\zeta\tau$ and denote $\bar{\zeta}=1-\zeta$.  Then the second term in (\ref{integral equation 1}) can be calculated by
\begin{eqnarray*}
&&\int^{\tau}_0G(\tau-s,z(\tau)-z(s))\phi(z(s))z'(s)ds\\
&=&\sum_{j=1}^JA_j\int_0^{\tau}\frac{1}{\sqrt{4\pi t}}e^{-\frac{[z(\tau)-z(\tau-t)-\kappa t]^2}{4t}-\rho t+q_jz(\tau-t)}z'(\tau-t)dt\\
&&{} -\nu K\int_0^{\tau}\frac{1}{\sqrt{4\pi t}}e^{-\frac{[z(\tau)-z(\tau-t)-\kappa t]^2}{4t}-\rho t+z(\tau-t)}z'(\tau-t)dt\\
&=&\sum_{j=1}^JA_j\sqrt{\tau}\int_0^{1}\frac{1}{\sqrt{4\pi \zeta}}e^{-\frac{[z(\tau)-z(\bar{\zeta}\tau)-\kappa \zeta\tau]^2}{4\zeta\tau} -\rho\zeta\tau+q_jz(\bar{\zeta}\tau)}z'(\bar{\zeta}\tau)d\zeta\\
&&{}-\nu K\sqrt{\tau}\int_0^{1}\frac{1}{\sqrt{4\pi \zeta}}e^{-\frac{[z(\tau)-z(\bar{\zeta}\tau)-\kappa \zeta\tau]^2}{4\zeta\tau} -\rho\zeta\tau+z(\bar{\zeta}\tau)}z'(\bar{\zeta}\tau)d\zeta\\
&:=& (\text{Term 1})+(\text{Term 2}).
\end{eqnarray*}
Using the expansions
\begin{eqnarray*}\label{eq-50}
&&z'(\bar{\zeta}\tau)=-A(\bar{\zeta}\tau)^{-1/2}(1+o(1)), \\
\label{eq-51}
&&e^{q_jz(\bar{\zeta}\tau)-\rho\zeta\tau}
=e^{q_jz_0}(1-2Aq_j\sqrt{\bar{\zeta}\tau}+o(\sqrt{\tau})),
\end{eqnarray*}
we derive that
\begin{eqnarray*}\label{eq-52}
(\text{Term 1})&=&\sum_{j=1}^JA_je^{q_jz_0}\tau^{\frac{1}{2}} \int_0^1\frac{1}{\sqrt{4\pi \zeta}}e^{-\frac{[z(\tau)-z(\bar{\zeta}\tau)-\kappa \zeta\tau]^2}{4\zeta\tau}} z'(\bar{\zeta}\tau)d\zeta\\
     &&{}-\sum_{j=1}^JAA_je^{q_jz_0}\int_0^1\frac{1}{\sqrt{4\pi \zeta}}e^{-\frac{[2A(\sqrt{\bar{\zeta}\tau}-\sqrt{\tau})+o(\sqrt{\tau})]^2} {4\zeta\tau}}\\
     &&\quad\cdot(1+o(1))\big(-2Aq_j\sqrt{\bar{\zeta}\tau}+o(\sqrt{\tau})\big) (\bar{\zeta})^{-1/2}d\zeta.
\end{eqnarray*}
Similarly, one can obtain that
\begin{eqnarray*}\label{eq-53}
(\text{Term 2}) &=&-\nu Ke^{z_0}\tau^{\frac{1}{2}}\int_0^1\frac{1}{\sqrt{4\pi \zeta}}e^{-\frac{[z(\tau)-z(\bar{\zeta}\tau)-\kappa \zeta\tau]^2} {4\zeta\tau}}z'(\bar{\zeta}\tau)d\zeta\\
     &&{}+A\nu Ke^{z_0}\int_0^1\frac{1}{\sqrt{4\pi \zeta}}e^{-\frac{[2A(\sqrt{\bar{\zeta}\tau}-\sqrt{\tau}) +o(\sqrt{\tau})]^2}{4\zeta\tau}}\\
     &&\quad\cdot (1+o(1))\left(-2A\sqrt{\bar{\zeta}\tau} +o(\sqrt{\tau})\right)(\bar{\zeta})^{-1/2}d\zeta.
\end{eqnarray*}
Since $\phi(z_0)=0$, this leads to
\begin{equation}\begin{split}\label{eq-54}
(\text{Term 1})+(\text{Term 2}) &=2A^2\left[\sum_{j=1}^Jq_jA_je^{q_jz_0}-\nu Ke^{z_0}\right]\\
&\quad \cdot (1+o(1))\sqrt{\tau}\int_0^1\frac{1}{\sqrt{4\pi \zeta}}e^{-\frac{[2A(\sqrt{\bar{\zeta}\tau}-\sqrt{\tau}) +o(\sqrt{\tau})]^2}{4\zeta\tau}}d\zeta+o(\sqrt{\tau}).
\end{split}\end{equation}
By (\ref{integral equation 1}), (\ref{eq-48}) and (\ref{eq-54}), we derive that
\begin{equation*}\label{eq-55}
\frac{1}{\sqrt{\pi}}e^{-A^2}-A\,\text{erfc}(A)=2A^2\int_0^1\frac{1}{\sqrt{4\pi \zeta}}e^{-\frac{[2A(\sqrt{\bar{\zeta}\tau}-\sqrt{\tau})]^2}{4\zeta\tau}}d\zeta.
\end{equation*}
By the transformation $\eta=\frac{1-\sqrt{\bar{\zeta}}}{\sqrt{\zeta}}A$, we  derive
\begin{equation}\label{transition-eqn}
\frac{1}{\sqrt{\pi}}e^{-A^2}-\frac{2A}{\sqrt{\pi}}\int_A^{\infty}e^{-\eta^2}d\eta=\frac{2}{\sqrt{\pi}}\int_0^Ae^{-\eta^2}\frac{A^3(A^2-\eta^2)}{(A^2+\eta^2)^2}d\eta.
\end{equation}
Let $F(A)=\frac{1}{\sqrt{\pi}A}e^{-A^2}- \text{erfc}(A)-\frac{2}{\sqrt{\pi}}{\displaystyle \int_0^{A}} e^{-\eta^2}\frac{A^2(A^2-\eta^2)}{(A^2+\eta^2)^2}d\eta$.
By a direct computation, we have
$$F'(A)=-\frac{1}{A^2\sqrt{\pi}}e^{-A^2}+\frac{2}{\sqrt{\pi}}\int_0^Ae^{-\eta^2} \frac{(-6A^3+2A\eta^2)\eta^2}{(A^2+\eta^2)^3}d\eta<0,$$
with $F(0)=+\infty$ and $F(+\infty)=-1$. This implies there exists a unique solution to the equation (\ref{transition-eqn}).
Finally, (\ref{eq-56-transedental}) follows from ${\displaystyle \int_A^{\infty}} e^{-\eta^2}d\eta=\frac{\sqrt{\pi}}{2}-{\displaystyle \int_0^A} e^{-\eta^2}d\eta$ and \eqref{transition-eqn}.
\endproof

 \subsection{Proof of Theorem \ref{Theorem 4.4}}
To prove Theorem \ref{Theorem 4.4}, we need the following lemma.
\begin{lemma}\label{var-ineq-inf-horizon}
There exists some $z^*\in\mathbb{R}$ such that
\begin{equation*}
\lim_{\tau\to\infty}z(\tau)=z^*.
\end{equation*}
\end{lemma}
\proof
Firstly, we consider the following problem
\begin{eqnarray}\label{ode-problem}
&&-\Psi''(z)+\kappa \Psi'(z)+\rho \Psi(z)=0\ \text{for}\ z>a,\nonumber\\
&&\Psi(z)=g(z)\ \text{for}\ z\leq a,\nonumber\\
&& \Psi'(a)=g'(a),\\
&&\lim_{z\to \infty}\Psi(z)=0. \nonumber
\end{eqnarray}
Denote
\begin{equation}\label{def-k}
p(z):=\sum_{j=1}^J-\frac{1}{q_j}(q_j-\lambda)e^{(q_j-1)z}-K(1-\lambda),
\end{equation}
where $\lambda=\frac{\kappa -\sqrt{\kappa^2+4\rho}}{2}$.
By Assumption \ref{parameter-assumption},   and $A_1<A_2<\ldots<A_J$ , we have $\kappa^2+4\rho-(\kappa-2q_j)^2=-4q_j A_j> 0$, which leads to
\begin{equation}\label{q-lambda}
q_j-\lambda=\frac{\sqrt{\kappa^2+4\rho}-\kappa+2q_j}{2}>0.
\end{equation}
This implies that $p'(z)<0$, ${\displaystyle\lim_{z\to\infty}} p(z)=-K(1-\lambda)<0$, and  ${\displaystyle\lim_{z\to-\infty}} p(z)=\infty$.
Hence, there exists a unique $a\in\mathbb{R}$ such that
\begin{equation}\label{def-z-star}
p(a)=0.
\end{equation}
Now the solution to problem \eqref{ode-problem} is given by
\begin{eqnarray}\label{solu-ode-problem}
&& \Psi(z)=g(a)e^{\lambda(z-a)}\ \text{for}\ z>a,\nonumber\\
&& \Psi(z)=g(z)\ \text{for}\ z\leq a,
\end{eqnarray}
where $a$ is defined by \eqref{def-z-star}.

We shall prove that the function $\Psi(z)$ defined by \eqref{solu-ode-problem} satisfies the following variational equation
\begin{equation}\label{var-ineq-inf-horizon-1}
\min\{-\Psi''+\kappa\Psi'+\rho\Psi,\, \Psi-g\}=0, \ z\in\mathbb{R}.
\end{equation}
 Firstly, for any $z>a,$  since $\Psi$ is the solution to problem \eqref{ode-problem}, we only need to verify $\Psi(z)>g(z)$.

Denote $\Phi(z,c)=g(c)e^{\lambda(z-c)}$. Differentiating $\Phi(z,c)$ in $c$ we have
$$\frac{\p}{\p c}\Phi(z,c)=e^{\lambda(z-c)+c}p(c),$$
where $p(c)$ is defined by \eqref{def-k}.
This implies that $\Phi(z,\cdot)$ is strictly increasing in $(-\infty,a)$ and strictly decreasing in $(a,z)$.
Hence, we have $\Psi(z)=\Phi(z,a)>\Phi(z,z)=g(z)$ for any $z> a.$ Consequently, $\Psi$ satisfies  \eqref{var-ineq-inf-horizon-1}
 for any $z>a$.

Secondly, for any $z\leq a,$ since $\phi(z_0)=\sum_{j=1}^JA_je^{q_jz_0}-K\nu e^{z_0}=0$, $\kappa=\nu-\rho+1$, $q_j-\lambda>0$ (see \eqref{q-lambda}), we have
\begin{eqnarray*}
\nu e^{z_0}p(z_0)&=&\nu e^{z_0}p(z_0)-\phi(z_0)\\
            &=&\sum_{j=1}^J e^{q_j  z_0}[-\frac{\nu}{q_j}(q_j-\lambda)-A_j]+\lambda K\nu e^{z_0}\\
&=& \sum_{j=1}^J e^{q_j  z_0}[-\frac{\nu}{q_j}(q_j-\lambda)+(\lambda-1)A_j] \\
&=&\sum_{j=1}^J-\frac{1}{4q_j} e^{q_j  z_0}(q_j-\lambda)[\kappa(2-2q_j)+(-2+2q_j)\sqrt{\kappa^2+4\rho}+4q_j-4]\\
&< & 0.
\end{eqnarray*}
Since $p(z)$ is strictly decreasing and $p(a)=0$, we derive that $z_0>a$. This leads to
\begin{equation*}
-g''+\kappa g'+\rho g=L[g]=\phi(z)>0 \ \text{for}\ z\leq a,
\end{equation*}
where the last inequality follows from the fact that $\phi$ is  strictly decreasing, $\phi(z_0)=0$, and $a<z_0$.
Thus $\Psi$ satisfies  \eqref{var-ineq-inf-horizon-1} for any $z\leq a$.

Now  the variational inequality  \eqref{var-ineq-inf-horizon-1}  implies that
\begin{eqnarray*}
&&\min\left\{L[\Psi], \ \Psi-g\right\}=0,\ (\tau,z)\in\Omega_T,\\
&&\Psi(z)\geq g(z)=v(0,z),\ z\in\mathbb{R}.
\end{eqnarray*}
By the comparison principle (see Lemma \ref{Lemma2.1}),
we have $v(\tau,z)\leq\Psi(z)$ for $(\tau,z)\in\Omega_T$.
Then  we derive that $z(\tau)\geq a$.  Otherwise, by the definition
of $z(\tau)$ (see \eqref{eq-24}) and \eqref{solu-ode-problem}, there exists some
$z\in (z(\tau),a)$ such that
$$v(\tau,z)>g(z)=\Psi(z).$$
The contradiction arises.
Since $z(\tau)$ is decreasing (See Theorem \ref{Theorem 3.1} ) and has a lower bound, there exists some $z^*\in\mathbb{R}$ such that
${\displaystyle\lim_{\tau\to\infty}}z(\tau)=z^*$.
\endproof

We can now prove Theorem \ref{Theorem 4.4}.
\proof We only need to show that $z^*=a$, where $a$ is defined in \eqref{def-z-star}.

We rewrite problem \eqref{eq-2} as
\begin{eqnarray*}
&&L[v]=I_{\{z\leq z(\tau)\}}\phi(z),\ (\tau,z)\in\Omega_T,\\
&&v(0,z)=g(z),\ z\in\mathbb{R},
\end{eqnarray*}
where  $I_A$ is the indicator function of set $A$.
By Green's identity, we have
\begin{equation*}\label{value-function-integral-1}
v(\tau,z)=\int_{-\infty}^{\infty}G(\tau,z-y)g(y)dy+\int_0^{\tau}\int_{-\infty}^{z(\tau-s)} G(s,z-y)\phi(y)dyds,
\end{equation*}
where $G$ is the Green function defined by (\ref{green}).
Since $v(\tau,z(\tau))=g(z(\tau))$ on the free boundary $(\tau,z(\tau))$, a  direct computation shows that
\begin{eqnarray*}\label{value-function-integral-2}
g(z(\tau))
&=& \int_{-\infty}^{\infty}G(\tau,z(\tau)-y)g(y)dy+\int_0^{\tau}\int_{-\infty}^{z(\tau-s)} G(s,z(\tau)-y)\phi(y)dyds\nonumber\\
&=&\sum_{j=1}^J-\frac{1}{q_j}\frac{1}{\sqrt{4\pi\tau}}e^{q_jz(\tau)+q_jA_j\tau}\int_{-\infty}^{\infty}\exp\Big[-\frac{(y-z(\tau)+\kappa\tau-2q_j\tau)^2}{4\tau}\Big]dy \nonumber\\
&&{}-K\frac{1}{\sqrt{4\pi\tau}}e^{z(\tau)-\nu\tau}\int_{-\infty}^{\infty}\exp\Big[-\frac{(y-z(\tau)+\kappa\tau-2\tau)^2}{4\tau}\Big]dy\nonumber\\
&&{}+ \sum_{j=1}^J A_j\int_0^{\tau}\frac{1}{\sqrt{4\pi s}}e^{q_jz(\tau)+q_jA_j s}\int_{-\infty}^{z(\tau-s)} \exp\Big[-\frac{(y-z(\tau)+\kappa s-2q_js)^2}{4s}\Big]dyds\nonumber\\
&&{}-\nu K \int_0^{\tau}\frac{1}{\sqrt{4\pi s}}e^{z(\tau)-\nu s}\int_{-\infty}^{z(\tau-s)} \exp\Big[-\frac{(y-z(\tau)+\kappa s-2s)^2}{4s}\Big]dyds \nonumber\\
&=&\sum_{j=1}^J-\frac{1}{q_j}e^{q_jz(\tau)+q_jA_j\tau}-K e^{z(\tau)-\nu\tau}\nonumber\\
&&{}+\sum_{j=1}^J A_j\int_0^{\tau}e^{q_jz(\tau)+q_jA_j s}
N\left(\frac{z(\tau-s)-z(\tau)+\kappa s-2q_j s}{\sqrt{2s}}\right)ds\\
&&{}-\nu K \int_0^{\tau}e^{z(\tau)-\nu s}
N\left(\frac{z(\tau-s)-z(\tau)+\kappa s-2 s}{\sqrt{2s}}\right) ds, \nonumber
\end{eqnarray*}
where $N(\cdot)$ is the cumulative distribution function of a standard normal variable.
Letting $\tau\to\infty$, by the dominated convergence theorem and the integration by parts, we have
\begin{eqnarray*}\label{value-function-integral-5}
g(z^*) &=&\sum_{j=1}^J A_j\int_0^{\infty}e^{q_j z^*+q_jA_j s}N\left(\frac{\kappa s-2q_j s}{\sqrt{2s}}\right)
ds-\nu K \int_0^{\infty}e^{z^*-\nu s}N\left(\frac{\kappa s-2 s}{\sqrt{2s}}\right) ds\\
 &=&\sum_{j=1}^J -\frac{1}{2q_j}e^{q_j z^*}\Big(1+\frac{\kappa-2q_j}{\sqrt{\kappa^2+4\rho}}\Big)
-\frac{1}{2}K e^{z_*}\Big(1+\frac{\kappa-2}{\sqrt{\kappa^2+4\rho}}\Big)\\
&=& \frac{1}{2}g(z^*)-\sum_{j=1}^J\frac{\kappa-2q_j}{2q_j\sqrt{\kappa^2+4\rho}}e^{q_j z^*}-\frac{\kappa-2}{2\sqrt{\kappa^2+4\rho}}Ke^{z^*}.
\end{eqnarray*}
Simple algebraic computation gives  $p(z^*)=0=p(a)$, where $p$ is defined in \eqref{def-k}. Hence, ${\displaystyle\lim_{\tau\to\infty}}z(\tau)=z^*=a$.
\endproof

 \subsection{Proof of Theorem \ref{Theorem 5.1}}
\proof Define the continuation region in $y$-coordinate to be
$\mathcal{C}_y=\{(t,y);\,\tilde V(t, y)>\tilde U_K(y),\, 0\leq t< T\}$,
and the exercise region to be
$\mathcal{S}_y=\{(t,y);\,\tilde V(t, y)=\tilde U_K(y),\, 0\leq t< T\}$.
The exercise boundary in $y$-coordinate is defined by
$y(t):=\inf\{y; \,\tilde V(t, y)>\tilde U_K(y)\}$ for $0\leq t< T$.
Then one can derive the global approximation of $y(t)$ by
\begin{equation*}\label{global approx 1}
y(t)\approx \exp\big(z_*(\tau)\big)=\exp\left(z_*\Big(\theta^2 (T-t)/2\Big)\right).
\end{equation*}
  From the dual transformation,
we know that
$\tilde{V}_y(t,y)=-x$.
On the free boundary, we have
$\tilde{V}_y(t,y(t))=\tilde{U}_K'(y(t))$.
Combining the above relations, we find the approximate free boundary
$x(t)$.

 From (\ref{eq-78}), also noting $u(0,y)=-\phi(y)$ and $u_z(s,z(s))=-\phi(z(s))z'(s)$, we have
\begin{eqnarray*}
v_{\tau}(\tau,z)&=&{}-\int_{z_0}^{\infty}G(\tau,z-y)\phi(y)dy+\int_0^{\tau}G(\tau-s,z-z(s))\phi(z(s))z'(s)ds\\
&=&{}-\int_{z_0}^{\infty}G(\tau,z-w)\phi(w)dw+\int_0^{\tau}G(\eta,z-z(\tau-\eta))\phi(z(\tau-\eta))z'(\tau-\eta)d\eta\\
&=&-\frac{\p}{\p\tau}\Big[\int_0^{\tau}\int_{z(\tau-\eta)}^{\infty}G(\eta, z-w)\phi(w)dwd\eta\Big].
\end{eqnarray*}
Integrating the above equation from $\tau=0$ to $\tau=\tau$ and noting $v(0,z)=g(z)$, we have
\begin{eqnarray*}
v(\tau,z)-g(z)&=& -\int_0^{\tau}\int_{z(\tau-\eta)}^{\infty}G(\eta, z-w)\phi(w)dwd\eta\\
                    &=& -\int_0^{\tau}\int_{z(s)}^{\infty}G(\tau-s, z-w)\phi(w)dwds,
\end{eqnarray*}
where $\phi$ is defined in \eqref{eqn-10}.
We approximate the free boundary  $z(\tau)$ by the global closed-form approximation $z_*(\tau)$ in (\ref{global approximation}) and get the approximation of the dual value function $\tilde{V}(t,y)$ as
(note $z=\log y$ and $\tau=\frac{\theta^2}{2}(T-t)$)
\begin{equation}\label{appro-1}
\tilde{V}(t,y) \approx \tilde{U}_K(y)-\int_0^{\tau} \int_{z_*(s)}^{\infty}G\left(\tau-s,\ln y-w\right)\phi(w)dwds.
\end{equation}
By Corollary \ref{cor2.3}, there exists a unique solution $y^*=I(t,x)$ to
the equation $\tilde{V}_y(t,y)+x=0$ for $x>K$. Then the primal value function is given by
$$V(t,x)=\inf_{y>0}(\tilde V(t,y)+xy)=\tilde{V}(t,I(t,x))+xI(t,x)$$
 by \eqref{appro-1} for any $(t,x)\in [0,T)\times[K,+\infty)$.
Finally, we  calculate the optimal strategy $\pi^*_t$. Since
$$V_x=y^*=I(t,x),\ V_{xx}(t,x)=-\frac{1}{\tilde{V}_{yy}(t,I(t,x))},$$
we derive that
$$\pi^*_t=-\frac{\theta}{\sigma}\frac{V_x}{V_{xx}}=\frac{\theta}{\sigma}I(t,x)\tilde{V}_{yy}(t,I(t,x)).$$
\endproof

\section{Conclusions}\label{sec:conclusion}

This paper provides a rigorous analysis of the optimal investment stopping problem using the dual control method. The analysis covers  a class of utility functions, including power and non-HARA utilities. The approximate formulas for the optimal value functions and optimal strategies are derived by developing the approximate formulas for the dual problems.
For non-HARA utility, if   Assumption \ref{parameter-assumption} does not hold, then there may exist two free boundaries or no free boundary for the dual problem   and we cannot use the method developed in this paper to characterize the limiting behaviour of the free boundary as time to maturity tends to zero or infinity, which makes impossible to find a global closed-form approximation to the free boundary.   We leave this for the future work.

\bigskip\noindent
{\bf Acknowledgments}. The authors are very grateful to  two anonymous reviewers whose constructive comments and suggestions have helped to improve the paper of the previous version.

\appendix
\section{Appendix: Proof of Theorem \ref{Theorem 4.1}}
 Theorem \ref{Theorem 4.1} is considered a known result in the PDE theory, but for the convenience of the reader, we give  a proof.
 Note that the payoff function for vanilla American option is Lipschitz continuous, but the function $g$ in
 (\ref{eq-4})
   is not Lipschitz continuous in the infinite region. So the analysis is different from that of  \cite{LIANG2007}.

Firstly, we  prove the following comparison principle:
\begin{lemma}\label{Lemma2.1}
Let $v_1,\, v_2\in W^{1,2}_{p,loc}(\Omega_{T})\cap C(\bar{\Omega}_{T})$ be functions satisfying
$|v_i|\leq C(e^{\alpha z}+e^{-\gamma z})$ for some positive  constants $C, \alpha, \gamma$, $i=1,2$,
and
\begin{eqnarray*}
&&F[v_1]\geq F[v_2], \ (\tau,z)\in \Omega_{T}, \\
&&v_1(0,z)\geq v_2(0,z),\ z\in \mathbb{R}^1,
\end{eqnarray*}
where $F[v]:=\min\left\{L[v], \ v-g\right\}$. Then
\begin{equation*}
v_{1}(\tau,z)\geq v_2(\tau,z), \ (\tau,z) \in \bar{\Omega}_{T}.
\end{equation*}
\end{lemma}
\proof
Note that on the set $\Omega_1:=\{(\tau,z)\in\Omega_T;\, v_2(\tau,z)-g(z)\leq L[v_2]\}$, we automatically have $v_1(\tau,z)-g(z)\geq F[v_1]\geq F[v_2]=v_2(\tau,z)-g(z)$,
so that $v_1(\tau,z)\geq v_2(\tau,z)$.
On the set $\Omega_2:=\{(\tau,z)\in\Omega_T;\, v_2(\tau,z)-g(z)> L[v_2]\}$, we have $L[v_1]\geq F[v_1]\geq F[v_2]=L[v_2]$.

We are now in a situation where
$L[v_1]\geq L[v_2]$ for $(\tau,z)\in\Omega_2$ and
$v_1(\tau,z)\geq v_2(\tau,z)$ for $(\tau,z)\in \bar{\Omega}_T-\Omega_2$.
We can apply the maximum principle (see \cite[Theorem 2.7]{LIEBERMAN1996}) on $\Omega_2$ to conclude that $v_1(\tau,z)\geq v_2(\tau,z)$ on $\Omega_2$.
\endproof

To prove the existence of the solution of problem (\ref{eq-2}), we construct a penalty function $\beta_{\epsilon}(t)\in C^2(\mathbb{R}^1)$  satisfying (see \cite{FRIEDMAN1982})
\begin{eqnarray*}
&&\beta_{\epsilon}(t)\leq 0, \ \beta_{\epsilon}(0)=-C_0 \ (C_0>0),\\
&&\beta_{\epsilon}(t)=0,\ t\geq\epsilon,\\
&&\beta'_{\epsilon}(t)\geq 0, \ \beta^{\prime\prime}_{\epsilon}(t)\leq 0,\\
&&\beta_{\epsilon}(t)\rightarrow 0,\quad \text{if}\; t>0,\ \epsilon\rightarrow 0,\\
&&\beta_{\epsilon}(t)\rightarrow -\infty,\quad \text{if}\; t<0,\ \epsilon\rightarrow 0,
\end{eqnarray*}
where $C_0$ is a constant to be determined.

Since system (\ref{eq-2})  lies in an unbounded domain, we apply a bounded domain to approximate it:
\begin{eqnarray}\label{eq-94}
&&\min\left\{L[v^{R}], \ v^{R}-g\right\}=0, \ (\tau,z)\in \Omega^R_T:=\left(0,\theta^2 T/2\right)\times(-R,R),\\
\label{eq-95}
&&v^{R}(\tau,z)=g(z),\ (\tau,z)\in \p_p\Omega^R_T,
\end{eqnarray}
where $\p_p\Omega^R_T$ is parabolic boundary, the operator $L$ and $g(z)$ is defined in \eqref{eq-4}.
Consider the penalty problem of \eqref{eq-94} - \eqref{eq-95}:
\begin{eqnarray}\label{eq-96}
&&L[v^{\epsilon,R}]+\beta_{\epsilon}(v^{\epsilon,R}-g)=0, \ (\tau,z) \in \Omega^R_{T},\\
\label{eq-97}
&&v^{\epsilon,R}(\tau,z)=g(z), \ (\tau, z)\in \p_p\Omega^R_T.
\end{eqnarray}
By \cite[Theorem 8.2]{FRIEDMAN1982}, For fixed $\epsilon$ and $R$, problem (\ref{eq-96}) - (\ref{eq-97}) has a unique solution $v=v^{\epsilon,R}\in W^{1,2}_p(\Omega^R_T)$, $1<p<+\infty$.

\begin{lemma}\label{Lemma 4.3}
For any fixed $R>0$, there exists a unique solution $v^R\in C(\bar{\Omega}^R_T)\cap W^{1,2}_p(\Omega^R_T)$ of problem (\ref{eq-94}) - (\ref{eq-95}), $1<p<+\infty. $ Moreover
\begin{equation}\label{eq-103}
g(z)\leq v^R(\tau,z)\leq \tilde{C}(e^{B\tau+\frac{p}{p-1}z}+1),\ (\tau,z)\in\Omega^R_T,
\end{equation}
where $\tilde{C}$ is defined as in  (\ref{im-1}), $B=|(\frac{p}{p-1})^2-\kappa \frac{p}{p-1}-\rho|+1$.
\end{lemma}
\proof
By \cite[Theorem 8.2]{FRIEDMAN1982}, we immediately obtain that there exists a unique  solution defined by $v^R:=\lim_{\epsilon\rightarrow0}v^{\epsilon,R}$ of the problem (\ref{eq-94}) - (\ref{eq-95}) and $v^R\in C(\bar{\Omega}^R_T)\cap W^{1,2}_p(\Omega^R_T)$. The variational inequality \eqref{eq-94} implies the first inequality in \eqref{eq-103}.

To obtain the second inequality in (\ref{eq-103}),
denote $w(\tau,z)=\tilde{C}(1+e^{B\tau+\frac{p}{p-1}z})$.
By (\ref{im-1}), we note that
$$w-g=w-\tilde{U}_K(e^z)\geq w-(\tilde{C}(1+e^{\frac{p}{p-1}z})-K e^z)\geq Ke^z\geq K e^{-R}\geq\epsilon$$
for small $\epsilon$ and $(\tau,z)\in\Omega^R_T$. By the definition of $\beta_{\epsilon}$,  this implies that
\begin{equation*}
\beta_{\epsilon}(w-g)=0.
\end{equation*}
Hence, by choosing $B=|(\frac{p}{p-1})^2-\kappa \frac{p}{p-1}-\rho|+1$, we have
\begin{equation*}\label{eq-105}
L[w]+\beta_{\epsilon}(w-g)=\tilde{C} e^{B\tau+\frac{p}{p-1}z}\left(B-\Big(\frac{p}{p-1}\Big)^2+\frac{p}{p-1}\kappa +\rho\right)+\tilde{C}\rho\geq 0.
\end{equation*}
The last inequality above follows from the definition of $A$ and $B$.
By the  comparison principle, we obtain
$$v^{\epsilon,R}\leq w \ \text{in}\ \Omega^R_T.$$
Now by letting $\epsilon\rightarrow 0$,  we complete the proof.
\endproof

We can now complete the proof of Theorem \ref{Theorem 4.1}.
\proof
By setting $R=n$ $(n\in\mathbb{Z}^+)$ in \eqref{eq-94} - \eqref{eq-95}, we rewrite the variational problem (\ref{eq-94}) - (\ref{eq-95}) as
\begin{eqnarray*}\label{eq-107}
&&L[v^n]=f(\tau,z),\ (\tau,z)\in\Omega_T^n,\\
\label{eq-108}
&&v^n(\tau,z)=g(z),\ z\in \p_p\Omega_T^n,
\end{eqnarray*}
with
$$f(\tau,z)=I_{\{v=g\}}L[g](z),$$
where $I_A$ is the indicator function of set $A$.
Combining (\ref{eq-103}), we deduce that for any fixed $\xi>0$,  the following $W^{1,2}_p$ interior estimate holds for $n>\xi$:
\begin{equation}\label{Lp interior estimates}
\|v^n\|_{W^{1,2}_p(\Omega^{\xi}_T)}\leq C_\xi,
\end{equation}
where  $C_\xi$  is a constant depending on $\xi$ but not on $n$, and $\|\cdot\|_{W^{1,2}_p(\Omega^{\xi}_T)}$ is the norm in the Sobolev space $W^{1,2}_p(\Omega^{\xi}_T)$.

Letting $\xi=1$ in \eqref{Lp interior estimates}. By the weak compactness and Sobolev embedding, there is a subsequence $\{v^n_{(1)}\}$ of $\{v^n\}$ such that
$$v^n_{(1)}\rightarrow v_{(1)} \ \text{weakly in} \ W^{1,2}_p(\Omega^1_T)$$
and
$$\|v^n_{(1)}-v_{(1)}\|_{C^0(\Omega^1_T)}\rightarrow 0.$$

Letting $\xi=2$ in \eqref{Lp interior estimates} with subsequence $\{v^n_{(1)}\}$ instead of $\{v^n\}$.
By the weak compactness and Sobolev embedding, there is a subsequence $\{v^n_{(2)}\}$ of $\{v^n_{(1)}\}$ such that
$$v^n_{(2)}\rightarrow v_{(2)} \ \text{weakly in} \ W^{1,2}_p(\Omega^2_T)$$
and
$$\|v^n_{(2)}-v_{(2)}\|_{C^0(\Omega^2_T)}\rightarrow 0.$$
Moreover, we  have
$$v_{(2)}=v_{(1)} \ \text{in}\ \Omega^1_T.$$
By induction, we conclude that there exists a subsequence $v^n_{(m)}$ of $v^n_{(m-1)}$ on $\Omega^m_T$ such that
$$v^n_{(m)}\rightarrow v_{(m)} \ \text{weakly in} \ W^{1,2}_p(\Omega^m_T)$$
and
$$\|v^n_{(m)}-v_{(m)}\|_{C^0(\Omega^m_T)}\rightarrow 0.$$
Moreover,
$$v_{(m)}=v_{(j)} \ \text{in}\ \Omega^j_T,\ 1\leq j\leq m-1.$$
We define $v=v_{(m)}$ if $(\tau,z)\in\Omega^m_T$ for any $m>0$. We consider the sequence $v^m_{(m)}$ in diagram. For any $N>0$,
 since $v^m_{(m)}$ is a subsequence of $v^m_{(N)}$ if $m>N$, we derive that
$$v^m_{(m)}\rightarrow v_{(N)}=v\ \text{weakly in} \ W^{1,2}_p(\Omega^N_T)$$
and
$$\|v^m_{(m)}-v\|_{C^0(\Omega^N_T)}=\|v^m_{(m)}-v_{(N)}\|_{C^0(\Omega^N_T)}\rightarrow 0.$$
Letting $m\rightarrow\infty$ in the system
\begin{eqnarray*}
&&\min\{L[v^m_{(m)}], \ v^m_{(m)}-g\}=0, \ (\tau,z)\in \Omega^m_T,\\
&&v^m_{(m)}(0,z)=g(z),\ z\in \p_p\Omega^m_T,
\end{eqnarray*}
we find that $v$ is the solution of problem (\ref{eq-2}).

The inequality (\ref{eq-106}) follows by letting $R\rightarrow\infty$ in the inequality (\ref{eq-103}).
 Lemma \ref{Lemma2.1} and (\ref{eq-106}) imply the uniqueness.

Finally, we prove (\ref{derivative-inequality}).
In the exercise region $\mathcal{S}_z$,  we  have
\begin{equation*}
v_z(\tau,z)=g'(z)=\tilde{U}_K'(e^z)e^z\leq 0
\mbox{ and }
-v_z(\tau,z)+v_{zz}(\tau,z)=\tilde{U}_K''(e^z)e^{2z}>0.
\end{equation*}
Note that the above inequalities also hold at time $\tau=0$ and at the boundary of  $\mathcal{C}_z$.
Since $L[v]=0$ in  $\mathcal{C}_z$, we have $L[v_z]=0$ and $L[-v_z+v_{zz}]=0$ for $(\tau,z)\in \mathcal{C}_z$.
The maximum principle implies that $ v_z\leq 0$ and $ - v_z+ v_{zz}> 0$ for $(\tau,z)\in \mathcal{C}_z$.

To prove $v_{\tau}\geq 0$,  we define
$$w(\tau,z)=v(\tau+\delta,z),\ \text{for small} \ \delta>0.$$
From (\ref{eq-2}), we know that $w(\tau,z)$ satisfies
\begin{eqnarray*}
&&\min\{L[w], \ w-g\}=0, \ (\tau,z) \in \tilde{\Omega}_T:=\left(0,\theta^2 T/2-\delta\right)\times\mathbb{R}^1,\\
&&w(0,z)=v(\delta,z)\geq g(z)=v(0,z),\ z\in \mathbb{R}^1.
\end{eqnarray*}
Applying the comparison principle (Lemma \ref{Lemma2.1}), we obtain that
$$w(\tau,z)=v(\tau+\delta,z)\geq v(\tau,z),\ \tau\in \left(0,\theta^2T/2-\delta\right),\ z\in\mathbb{R}^1.$$
Thus we have $v_{\tau}\geq 0$.
\endproof

\end{document}